\documentclass[%
 reprint,
 superscriptaddress,
 amsmath,amssymb,
 aps,
 pre,
 floatfix,
]{revtex4-2}

\usepackage{graphicx}
\usepackage{dcolumn}
\usepackage{bm}
\usepackage{hyperref}
\usepackage[dvipsnames]{xcolor}
\hypersetup{
    colorlinks = true,
    linkcolor = blue,
    anchorcolor = blue,
    citecolor = blue,
    filecolor = blue,
    urlcolor = black
}

\newcommand{\added}[1]{#1}
\usepackage{float}

\begin{document}

\preprint{APS/123-QED}

\title{Phase-Field Model of Freeze Casting}

\author{Kaihua Ji}
 \email[]{ji5@llnl.gov}
\affiliation{%
Department of Physics and Center for Interdisciplinary Research on Complex Systems, Northeastern University, Boston, Massachusetts 02115, USA
}%
\affiliation{%
Lawrence Livermore National Laboratory, Livermore, 94550, CA, USA
}%
\author{Alain Karma}%
 \email[]{a.karma@northeastern.edu; Corresponding author}
\affiliation{%
Department of Physics and Center for Interdisciplinary Research on Complex Systems, Northeastern University, Boston, Massachusetts 02115, USA
}%

\date{\today}

\begin{abstract}
Directional solidification of water-based solutions has emerged as a versatile technique for templating hierarchical porous materials. However, the underlying mechanisms of pattern formation remain incompletely understood. In this work, we present a detailed derivation and analysis of a quantitative phase-field model for simulating this nonequilibrium process.
The phase-field model extends the thin-interface formulation of dilute binary alloy solidification with anti-trapping to incorporate the highly anisotropic energetic and kinetic properties of the partially faceted ice-water interface. This interface is faceted in the basal plane normal to the $\left<0001\right>$ directions and atomically rough in other directions within the basal plane. On the basal plane, the model reproduces a linear or nonlinear relationship between the interface growth rate and the kinetic undercooling that can be linked to experimental measurements. In both cases, spontaneous parity breaking of the solidification front is observed when the preferred growth direction is aligned with the temperature gradient. This phenomenon leads to the formation of partially faceted ice lamellae that drift laterally in one of the $\left<0001\right>$ directions. We demonstrate that the drifting velocity of the ice lamellae is controlled by the kinetics on the basal plane and converges as the thickness of the diffuse solid-liquid interface decreases.
Furthermore, we examine the effect of the form of the kinetic anisotropy, which is chosen here such that the inverse of the kinetic coefficient varies linearly from a finite value in the $\left<0001\right>$ directions to zero in all other directions within the basal plane, consistent with the assumption that the interface grows in local thermodynamic equilibrium in this plane. Our results indicate that the drifting velocity of ice lamellae is not affected by the slope of this linear relation, and the radius and undercooling at the tip of an ice lamella converge at relatively small slope values. Consequently, the phase-field simulations remain quantitative with computationally tractable choices of both the interface thickness and the slope assumed in the form of the kinetic anisotropy.
\end{abstract}

\maketitle


\section{Introduction}

Freeze casting, also known as ice templating when the solvent is water, is a directional solidification technique that utilizes growing crystals to template a secondary phase, thereby producing hierarchical porous materials with tunable structures and properties \cite{Deville2006FreezingComposites,Wegst2010BiomaterialsCasting,Donius2014SuperiorCasting,scotti2018freeze,wegst2024freeze}. In this process, dissolved solutes or suspended particles in water-based solutions or slurries are segregated between advancing ice lamellae, forming hierarchical architectures that remain after the ice phase is removed via sublimation. These hierarchical architectures typically feature lamellar cell walls with unilateral surface features, ranging from secondary ridges perpendicular to the primary ice lamellae to more exotic patterns resembling living forms \cite{Yin2023HierarchicalTemplating}. Due to their hierarchical structures and controllable features, ice-templated materials have shown potential in various applications, ranging from biomedicine \cite{Riblett2012Ice-TemplatedGrowth,Bozkurt2012TheNerves,Deville2006FreezeEngineering,Mohan2018FluorescentModel,yin2019freeze} to energy generation and storage \cite{Han2019AnisotropicAbsorption,Qiu2022ExcellentFoams,Shao2020FreezeApplications}.

Despite this technological appeal, the mechanisms by which faceted ice crystals evolve to form such hierarchical architectures remain incompletely understood. In contrast to metallurgical alloys, which typically exhibit atomically rough, weakly anisotropic interfaces that are well-described by existing theories of cellular and dendritic growth, ice crystals exhibit strongly anisotropic growth behaviors \cite{Woodruff1973TheInterface,Libbrecht2017}. Growth along the slow $c$-axis of ice crystals (the $\left[0001\right]$ growth direction normal to the basal plane) is faceted and far from equilibrium, being primarily controlled by interface kinetics. Conversely, growth within the basal plane occurs approximately in local equilibrium and is controlled by solutal and thermal transport. This interplay of faceted and non-faceted regions on the ice-water interface drives the development of intricate microstructural patterns, such as the complex unilateral features observed in freeze-cast materials \cite{Yin2023HierarchicalTemplating}. To fully understand the formation of these microstructures, a quantitative computational model is needed to bridge the highly anisotropic interface dynamics with the resulting microstructural pattern formation.

The phase-field (PF) method is a powerful and versatile computational approach for simulating free-boundary problems without explicitly tracking interfaces \cite{Boettinger2002Phase-fieldSolidification,Chen2002Phase-fieldEvolution,steinbach_phase-field_2009,tourret_phase-field_2022}. This method has provided quantitative insights into the microstructural pattern formation observed in non-faceted systems, such as the microstructure selection during near-equilibrium \cite{Haxhimali2006OrientationEvolution,Bergeon2013SpatiotemporalSolidification,Clarke2017MicrostructureSimulations} and far-from-equilibrium \cite{ji_microstructural_2023,ji2024microstructure,ji2024quantitative} alloy solidification. However, its application to systems with strongly anisotropic and partially faceted interfaces remains comparatively limited, particularly in the context of ice-water solidification. The PF method has been used to study some faceted systems, including the solidification of cubic-symmetry faceted crystals \cite{Debierre2003Phase-fieldSolidification} and the complex dynamics of snowflake growth from supersaturated water vapor \cite{Demange2017ADimensions,Demange2017GrowthModel}. While it has also been applied to the freeze-casting process \cite{huang2018phase,Seiz2021ModellingMethod}, existing studies are limited to two dimensions (2D) and not able to capture the characteristic ice-templated structures observed experimentally. A significant challenge in quantitative modeling lies in accurately capturing the distinct kinetic behaviors of faceted versus non-faceted orientations on the same crystal: while basal plane directions grow approximately in local equilibrium with weakly anisotropic excess free energy, growth along $c$-axis is faceted and far from equilibrium, and this requires the implementation of highly anisotropic interface properties.

Here, we develop a quantitative PF model that captures the highly anisotropic faceted growth by smoothly interpolating between kinetically distinct regimes on the ice-water interface using anisotropy functions of interface orientations. This model was used in a recent study \cite{Yin2023HierarchicalTemplating} that combined simulations and experiments of binary water-sugar and ternary chitosan-acetic-acid-water systems. The results revealed that by localizing diffusion-limited morphological instabilities to the atomically rough interfaces of partially faceted lamellae, anisotropic ice-crystal growth templates hierarchical architectures. 
Characteristic structures, such as unilateral lamellae with subfeatures on the rough side, were reproduced quantitatively in simulations of simple binary water-based systems that resemble the complex hierarchical structures observed in more chemically complex freeze-casting systems. PF simulations further revealed a scaling law for the lamellar spacing, $\lambda \sim(V G)^{-1 / 2}$, where $V$ and $G$ are the local growth rate and temperature gradient, respectively. 
The same PF model used in Refs.~\cite{Yin2023HierarchicalTemplating,ji2021phase} was also applied in a recent study \cite{chen2025asymmetric} to simulate the asymmetric ice crystals observed in freeze casting of dilute water-NaCl solutions.

In this paper, we present a detailed exposition of the quantitative PF model for freeze casting including a study of its convergence properties; in a companion Letter \cite{Ji2025ParityTemplating}, we further carried out numerical simulations of this PF model to understand the dynamical selection of the growth orientation of lamellae within the theoretical framework of parity breaking. Here, we also investigate the interplay between weakly anisotropic excess interface free energy and highly anisotropic attachment kinetics. The anisotropy of the excess interface free energy has two cusps along the $\left<0001\right>$ directions, which result in small facets in the equilibrium shape obtained through the Wulff construction in three dimensions (3D). However, this is insufficient to reproduce the experimentally observed large facets in the lamellar structure, and coupling to kinetic anisotropy is required. 
On the basal plane, the PF model reproduces either a linear or nonlinear relationship between the interface growth rate and the kinetic undercooling. While experimental measurements indicate that basal plane growth kinetics is generally nonlinear over a large range of undercooling \cite{Libbrecht2017}, reflecting layer-by-layer growth mechanisms, these kinetics can be approximated by a linear relationship for the limited range of undercooling relevant to the narrow tip region that controls ice lamellar growth.
The separate effects of interface free-energy anisotropy and kinetic anisotropy on structural formation are then investigated. 3D simulations indicate that the highly anisotropic interface kinetics is crucial for spontaneous symmetry breaking, which drives the formation of partially faceted ice lamellae. While weakly anisotropic interface free energy cannot independently induce the formation of lamellar structures, it controls the solidification front and selects steady-state ice tips, which dictate the formation of unilateral surface features on the rough side of the lamellae. Quantitative PF simulations require the accurate implementation of both anisotropies.

We examine the convergence of the model as a function of the diffuse-interface thickness $W$ and the slope $r$, which characterizes the form of the kinetic anisotropy. The slope $r$ determines the rate of variation of the inverse of the kinetic coefficient, transitioning from a finite value in the $\left<0001\right>$ directions to vanishing in other directions within the basal plane where growth is assumed to take place with local equilibrium at the solid-liquid interface. We make use of the anti-trapping formulation of dilute binary alloy solidification \cite{Karma2001Phase-fieldSolidification,Echebarria2004QuantitativeSolidification} to model the limit of local equilibrium at the interface for growth within this plane, and introduce an anisotropic form of the relaxation time of the phase-field kinetics to smoothly interpolate between this local-equilibrium growth mode in the basal plane and the kinetically dominated growth mode on facets. Convergence tests for $W$ and $r$ are non-trivial due to the presence of connected faceted and non-faceted regions on the solid-liquid interface, which exhibit significantly different growth kinetics. We evaluated the drifting velocity of partially faceted ice lamellae, the ice tip radius, and the tip undercooling as functions of $W$ and $r$. Our results indicate that these parameters can be chosen to ensure computational tractability while maintaining quantitative accuracy. Additionally, we performed a detailed analysis of the kinetics on the facet below the solidification front. Our results show that the dynamics of partially faceted ice lamellae are primarily influenced by basal plane kinetics, and the lateral drifting is entirely controlled by this kinetics for $W$ smaller than a critical value, where the drifting dynamics becomes converged.

The paper is organized as follows. Section~\ref{Sec:sharp_interface} describes the sharp-interface equations governing the directional solidification of dilute binary water-based solutions. Section~\ref{Sec:phase_field_model} presents the derivation of the PF model in the complete-partitioning limit. Section~\ref{sec:anisotropies} focuses on the implementation of interface free-energy and kinetic anisotropies. Section~\ref{sec:numerical_results} presents the simulation results and additional analyses.

\section{Sharp interface equations} \label{Sec:sharp_interface}

We consider the solidification of a binary aqueous solution with a straight liquidus having a slope $m$ in the dilute range. The partition coefficient $k = c_s / c_l$ is close to zero, as nearly all solute is rejected by the growing ice phase during freezing at relatively low velocities. Here, $c_s$ and $c_l$ are the solute concentrations at the solid and liquid sides of the interface, respectively. For a moving solid-liquid interface with a normal velocity $V_n$, its temperature satisfies the generalized Gibbs-Thomson relation:
\begin{equation}
T = T_M - |m| c_l - \Gamma_{\mathrm{GT}} \mathcal{K} - {V_n}/{\mu_k}, \label{Gibbs}
\end{equation}
where $T_M$ is the melting temperature of pure ice, $\Gamma_{\mathrm{GT}} = \Gamma T_M / \Delta h_f$ is the Gibbs-Thomson constant, $\Gamma$ is the excess interfacial free energy, $\Delta h_f$ is the latent heat of fusion per unit volume, $\mathcal{K}$ is the interface curvature, and $\mu_k$ is the atomic attachment kinetic coefficient.

\added{We consider the limit of complete partitioning (i.e., $k \to 0$) for ice  crystal growth at low interface velocities under atmospheric pressure~\cite{tsironi2020brine}. This assumption may deviate slightly at high interface velocities~\cite{Peppin2008ExperimentalSuspensions} or under high pressure~\cite{journaux2017salt}, but such conditions are beyond the scope of this paper.} Under complete partitioning, solute molecules diffuse only on the liquid side of the interface, resulting in a one-sided model of solidification. For dilute impurities or small particles, diffusion follows the Fickian model and is much slower than the diffusion of latent heat released at the interface due to the liquid-solid phase transformation. By considering the temperature field fixed by external conditions and neglecting latent heat diffusion, the boundary condition at the interface can be derived from Eq.~\eqref{Gibbs}. In the classical sharp-interface model of solidification, the solid and liquid phases are separated by a sharp boundary, governed by the following equations describing the solidification dynamics:
\begin{gather}
\partial_t c = D \nabla^2 c, \label{sharp1} \\
c_l V_n = -\left. D \partial_n c \right|^+, \label{sharp2} \\
{c_l}/{c_\infty} = {c_l^0(T)}/{c_\infty} - d_0 \mathcal{K} - \beta_k V_n. \label{sharp3}
\end{gather}
Eq.~\eqref{sharp1} represents Fickian diffusion, where $c$ is the solute concentration in mole fraction and $D$ is the solute diffusivity in the liquid. Eq.~\eqref{sharp2} is the Stefan condition for mass conservation at the interface, where $\left. \partial_n c \right|^+$ is the derivative of $c$ on the liquid side of the interface. Eq.~\eqref{sharp3} is derived from Eq.~\eqref{Gibbs}, where $c_\infty$ is the nominal sample concentration, $c_l^0(T)$ is the equilibrium liquidus concentration at temperature $T$ (satisfying $T \leq T_M$), and
\begin{equation}
d_0 = \frac{\Gamma}{\Delta T_0}, \label{capillary_Delta_T}
\end{equation}
is the capillary length, with $\Delta T_0 \equiv |m| c_\infty$. Finally, $\beta_k = 1 / (\mu_k \Delta T_0)$ is defined as the reciprocal kinetic coefficient.

For directional solidification in a temperature gradient $G$ with a pulling/isotherm velocity $V_p$, the frozen temperature approximation defines the temperature field along the vertical $x$-axis as:
\begin{equation}
T(x) = T_0 + G\left(x - V_p t\right),
\end{equation}
where $T_0$ is chosen as the liquidus temperature $T_0 = T_M - |m| c_\infty$. With this thermal condition, the scaled liquidus concentration $c_l^0(T) / c_\infty$ in Eq.~\eqref{sharp3} (denoted hereafter by $\tilde{c}_l^0(T)$)  becomes:
\begin{equation}
\tilde{c}_l^0(T) = 1 - \frac{x - V_p t}{l_T}, \label{directional_c}
\end{equation}
with the thermal length defined as:
\begin{equation}
l_T = \frac{|m| c_\infty}{G}.
\end{equation}

\section{Quantitative phase-field model with complete partitioning} \label{Sec:phase_field_model}

\subsection{Variational formulation} \label{Sec:Variational}

We consider a PF model for the solidification of binary liquid mixtures of $A$ (water) and $B$ (solute). In this model, a scalar phase field $\phi$ takes on constant values in the solid ($\phi = +1$) and liquid ($\phi = -1$) phases, and varies smoothly across the diffuse interface. Another scalar field $c$ denotes the solute concentration, defined as the mole fraction of $B$. The two-phase system is described by a phenomenological free-energy functional \cite{karma_phase-field_2003},
\begin{equation}
F[\phi, c, T] = \int_{dV} \left[ \frac{\sigma}{2} |\vec{\nabla} \phi|^{2} + f\left(\phi, T_M\right) + f_{AB}(\phi, c, T) \right]. \label{F_functional}
\end{equation}
Inside the integrand, the first term is the gradient energy, ensuring a finite interface thickness; the second term is a double-well potential that stabilizes the two phases $\phi = \pm 1$, expressed as
\begin{equation}
f\left(\phi, T_M\right) = h\left(-\frac{\phi^{2}}{2} + \frac{\phi^{4}}{4}\right),
\end{equation}
with $h$ representing the barrier height; the third term, $f_{AB}(\phi, c, T)$, is the bulk free-energy density of the binary mixture. 
In the dilute limit, $f_{AB}(\phi, c, T)$ can be expressed as
\begin{equation}
f_{AB}(\phi, c, T) = f_A(\phi, T) + \frac{R_0 T_M}{v_0} \left( c \ln c - c \right) + \epsilon(\phi) c, \label{f_AB}
\end{equation}
where $f_A(\phi, T)$ denotes the free energy of pure $A$. The second term on the right-hand side of Eq.~\eqref{f_AB} is the standard entropy of mixing for dilute solutions, where $R_0$ is the gas constant, and $v_0$ is the constant molar volume. The third term accounts for the enthalpy of mixing, with $\epsilon(\phi)$ being an interpolation function. 
A possible choice for $\epsilon(\phi)$ is
\begin{equation}
\epsilon(\phi) = -\epsilon_0 \ln \left[ \frac{1 - g(\phi) + \delta}{2} \right],
\end{equation}
where $\epsilon_0$ is a constant and $\delta \ll 1$. The function $g(\phi)$ interpolates the enthalpy of mixing between the solid and liquid phases, satisfying $g(\pm 1) = \pm 1$ and $g'(\pm 1) = 0$. We choose its form to be
\begin{equation}
g(\phi)=\phi-\frac{2\phi^3}{3}+\frac{\phi^5}{5}. \label{interpolation_g}
\end{equation}
This choice ensures $\epsilon(-1) \to 0$ and $\epsilon(+1) \to \infty$ as $\delta \to 0$, indicating that the binary mixture behaves as an ideal solution in the liquid phase while mixing in the solid phase incurs an extremely large energy penalty, effectively insuring complete partitioning of the solute.

The system evolves to minimize the free energy. The dynamical evolution of $\phi$ and $c$ follows standard variational forms for non-conserved and conserved dynamics, respectively:
\begin{align}
\frac{\partial \phi}{\partial t} &= -K \frac{\delta F}{\delta \phi}, \label{eq_phi} \\
\frac{\partial c}{\partial t} &= \vec{\nabla} \cdot \left( M(c) \vec{\nabla} \frac{\delta F}{\delta c} \right), \label{eq_c}
\end{align}
where $K$ is a coefficient related to interface kinetics, and $M(c)$ is the mobility of solute atoms or molecules. We set $M(c) = M_0 c$ to restore Fickian diffusion in the liquid. Since complete partitioning is assumed, an interpolation function for the diffusion coefficient, as used in alloy solidification, is not required, and $M_0$ can be treated as a constant.
At equilibrium, $\partial_t \phi = \partial_t c = 0$, and Eqs.~\eqref{eq_phi}–\eqref{eq_c} reduce to:
\begin{align}
\frac{\delta F}{\delta \phi} &= 0, \label{eq_phi_equil} \\
\frac{\delta F}{\delta c} &= \mu_E(T), \label{eq_c_equil}
\end{align}
where $\mu_E(T)$ is the spatially uniform equilibrium value of the chemical potential, which depends on the temperature $T$ at the interface.

According to the equilibrium condition in Eq.~\eqref{eq_c_equil}, the chemical potential is spatially uniform throughout the entire two-phase system. This implies:
\begin{equation}
\left.\frac{\partial f_{AB}(\phi, c, T)}{\partial c}\right|_{c=c_0} = \left.\frac{\partial f_{AB}(-1, c, T)}{\partial c}\right|_{c=c_l^0} = \mu_{E}(T), \label{equilibrium_1}
\end{equation}
where $c_0$ is the equilibrium concentration profile across the diffuse interface, and $c_l^0$ is the concentration on the liquid side of the interface at equilibrium. From Eq.~\eqref{equilibrium_1}, the concentration profile $c_0(\phi)$ is given by:
\begin{equation}
c_0(\phi) = c_l^0 \exp\left[\frac{\epsilon_0 v_0}{R_0 T_M} \ln\left(\frac{1 - g(\phi) + \delta}{2}\right)\right].
\end{equation}
By choosing $\epsilon_0 = R_0 T_M / v_0$, the concentration profile simplifies to:
\begin{equation}
c_0(\phi) = c_l^0 \frac{1 - g(\phi) + \delta}{2}.
\end{equation}
Using the equilibrium condition from Eq.~\eqref{eq_phi_equil}, we obtain:
\begin{equation}
f_A(\phi, T) = -\frac{\epsilon_0}{2} g(\phi) c_l^0(T),
\end{equation}
where $c_l^0(T)$ is determined by the liquidus in the phase diagram.

Thus, we obtain a complete expression for $f_{AB}(\phi, c, T)$ that contains the interpolation function $g(\phi)$. By substituting Eq.~\eqref{F_functional} into Eqs.~\eqref{eq_phi}–\eqref{eq_c}, and defining the constants $\tau = 1 / Kh$, $W = \sqrt{\sigma / h}$, $D = M_0 R_0 T_M / v_0$, and $\lambda = \epsilon_0 c_\infty / 2h$, the equations of motion become:
\begin{gather}
\tau \partial_t \phi = W^2 \nabla^2 \phi + \phi - \phi^3 - \lambda g'(\phi) \left[e^u - \tilde{c}_l^0(T)\right], \\
\partial_t c = \vec{\nabla} \cdot \left(D c \vec{\nabla} u\right), \label{eom_c_var}
\end{gather}
with
\begin{equation}
u = \ln\left(\frac{2c / c_\infty}{1 - g(\phi) + \delta}\right). \label{chem_devi_ug}
\end{equation}
The variable $u$ represents a dimensionless measure of the deviation of the chemical potential from its equilibrium value for a flat interface, where the liquid-side concentration equals $c_\infty$ at a fixed reference liquidus temperature $T_0 = T_M - |m| c_\infty$. 

\paragraph*{Limitation of the Variational Model.}

In the quantitative PF model for alloy solidification \cite{Karma2001Phase-fieldSolidification,Echebarria2004QuantitativeSolidification}, two independent interpolation functions, $g(\phi)$ and $q(\phi)$, are introduced within the variational formulation. The function $g(\phi)$ [see Eq.~\eqref{interpolation_g}] interpolates the enthalpy of mixing between the solid and liquid phases, while $q(\phi)$ [see Eq.~\eqref{interpolation_q}] is used to interpolate the solute diffusivity across the diffuse interface in a one-sided model. However, these two functions alone are insufficient to satisfy all three constraints required to cancel excess quantities---including surface diffusion, interface stretching, and the chemical potential jump at the interface---that arise due to the use of a mesoscopic diffuse interface thickness $W$ in the PF model.
To address this limitation, a phenomenological ``anti-trapping'' current was introduced in a nonvariational formulation \cite{Karma2001Phase-fieldSolidification}. This provides an additional degree of freedom that can be utilized to satisfy all three constraints, ensuring that the PF model remains quantitative for a mesoscopic $W$.
In comparison, the present variational PF model for ice templating employs only one interpolation function $g(\phi)$. The function $q(\phi)$ is not needed in the limit of complete partitioning. The function $g(\phi)$ can be chosen to meet the first two constraints (Eqs.~(4)-(5) in Ref.~\cite{Echebarria2004QuantitativeSolidification}), which correspond to excess surface diffusion and interface stretching that reduce to a single constraint in the limit of complete partitioning. However, to eliminate the discontinuity in chemical potential at the interface and control the interface kinetics, an additional degree of freedom is necessary. This indicates a limitation of the present variational formulation.
Therefore, we adopt a nonvariational formulation by including an anti-trapping current, which is necessary for controlling the interface kinetics of a highly anisotropic ice-water interface.

\subsection{Nonvariational formulation}

For modeling alloy solidification with a finite partitioning coefficient, an anti-trapping current $\vec{j}_{at}$ is introduced into the PF model to compensate for the spurious solute trapping caused by a mesoscopic interface thickness $W$ \cite{Karma2001Phase-fieldSolidification,Echebarria2004QuantitativeSolidification}. This provides additional freedom to achieve the correct mapping between the diffuse interface model and the desired free-boundary problem. By selecting the form of $\vec{j}_{at}$ through asymptotic analysis, the chemical potential jump at the interface can be eliminated, ensuring that the interface growth occurs under local equilibrium conditions. 
For quantitative modeling of ice-crystal growth, the additional freedom provided by $\vec{j}_{at}$ is utilized to precisely control the interface driving force. Combined with a carefully chosen form of the relaxation time $\tau$, as presented subsequently, it is possible to restore local equilibrium 
for fast growth parallel to the basal plane, while maintaining out-of-equilibrium conditions 
for faceted growth normal to the basal plane. Since solute trapping vanishes in the limit of complete partitioning, the physical interpretation of the ``anti-trapping'' current is slightly different in the PF model for ice templating: it redistributes solute across the diffuse interface with a mesoscopic $W$, thereby eliminating the discontinuity in chemical potential. This formulation retains a degree of freedom to vary the relaxation time $\tau$ to recover the desired kinetic undercooling, which can be either vanishing or finite depending on the interface orientation.

With the incorporation of the anti-trapping current $\vec{j}_{at}$, the continuity relation in Eq.~\eqref{eom_c_var} becomes:
\begin{equation}
\partial_{t} c = \vec{\nabla} \cdot \left( D c \vec{\nabla} u - \vec{j}_{at} \right), \label{eom_c_at}
\end{equation}
where
\begin{equation}
\vec{j}_{at} = - a W c_{\infty} e^u \frac{\partial \phi}{\partial t} \frac{\vec{\nabla} \phi}{|\vec{\nabla} \phi|}.
\end{equation}
Since the condition $g'(\pm 1) = 0$ is not required in the continuity equation, the function $g(\phi)$ in Eq.~\eqref{chem_devi_ug} can be replaced by $h(\phi)$, which satisfies $h(\pm 1) = \pm 1$ but does not necessarily satisfy $h'(\pm 1) = 0$. The expression for $u$ in Eq.~\eqref{eom_c_at} is then:
\begin{equation}
u = \ln\left(\frac{2c / c_\infty}{1 - h(\phi) + \delta}\right). \label{chem_devi_uh}
\end{equation}
We select $h(\phi) = \phi$ to ensure that the equilibrium concentration profile across the diffuse interface has the lowest possible gradients \cite{Echebarria2004QuantitativeSolidification}.

The PF model with complete partitioning (referred to herein as the \textit{$\delta$-model}) and the PF model with finite partitioning (referred to herein as the \textit{$k$-model}, with details provided in Appendix \ref{sec:k-model}) solves quantitatively the same sharp-interface equations in the limit $k \to 0$. 
The $k$-model and $\delta$-model differ primarily in the mathematical form used to regularize the governing equations near the solid phase ($\phi \to 1$): the $k$-model employs a small finite partition coefficient $k$, while the $\delta$-model introduces a regularization parameter $\delta$ to achieve complete partitioning. In both models, these parameters serve the same purpose of preventing the divergence of the enthalpy of solute addition in the solid in the limit of vanishing $\delta$ or $k$. As a result, the two models are effectively equivalent in their asymptotic limits, and their quantitative predictions are expected to be similar. The main distinction lies in the choice of regularization form, which has only a minor effect on simulation outcomes. Due to this effective equivalence, either model can be employed to simulate ice templating, with the $\delta$-model offering a slight computational advantage due to requiring one fewer interpolation function.

The asymptotic analysis presented in Ref.~\cite{Echebarria2004QuantitativeSolidification} for the PF model with finite partitioning is also applicable to the $\delta$-model discussed here. With the choices of $g(\phi)$ in Eq.~\eqref{interpolation_g} and $a = 1 / (2\sqrt{2})$, both following Ref.~\cite{Echebarria2004QuantitativeSolidification}, it is straightforward to show that the sharp-interface equations in Eqs.~\eqref{sharp1}–\eqref{sharp3} are recovered by the present PF model in its thin-interface limit.
The final result of the asymptotic analysis yields a kinetic relation, with the reciprocal kinetic coefficient $\beta_k$ in Eq.~\eqref{sharp3} expressed in terms of the PF model parameters as:
\begin{equation}
\beta_k(T, \mathbf{n}) = \frac{a_1 \tau(T, \mathbf{n})}{\lambda W(\mathbf{n})} - \frac{a_1 a_2 W(\mathbf{n})}{D} \tilde{c}_l^0(T). \label{beta}
\end{equation}
Both $W$ and $\tau$ are orientation-dependent due to the anisotropic properties of the interface. The interface thickness is given by $W(\mathbf{n}) = W_0 a_s(\mathbf{n})$, where the anisotropy function $a_s(\mathbf{n})$ for the interface free energy is introduced in Sec.~\ref{Sec:free_energy}. As discussed in Sec.~\ref{Sec:kinetic_anisotropy}, the relaxation time $\tau(T, \mathbf{n})$ is chosen as a function of both temperature and orientation. This choice ensures that $\beta_k$ vanishes for growth directions perpendicular to the $c$-axis within the basal plane, while $\beta_k$ takes a finite value along the $c$-axis.

\section{Anisotropic ice-water interface} \label{sec:anisotropies}

The accurate simulation of the ice-water interface requires the implementation of both a weakly anisotropic excess free energy and a highly anisotropic interface kinetics. In this section, we first discuss the anisotropy in excess free energy, focusing on the cusps along the $\left<0001\right>$ directions and the resulting equilibrium shape. Next, we examine the anisotropy of atomic attachment kinetics, and consider both linear and nonlinear kinetic relationships on the basal plane.

\subsection{Anisotropy of excess interface free-energy} \label{Sec:free_energy}

\begin{figure*}
\includegraphics[scale=1]{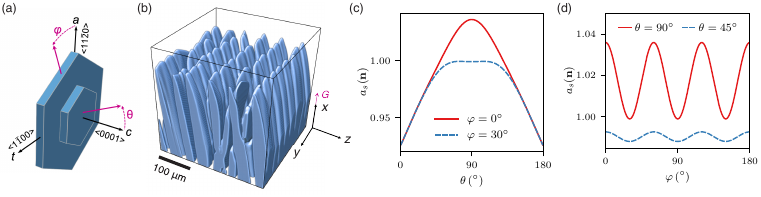}
\caption{(a) Hexagonal ice crystal. (b) Ice phase in a 3D PF simulation of the directional solidification of a 3 wt.\% aqueous sucrose solution with growth conditions $V_p = 15~\mathrm{\mu m/s}$ and $G = 12~\mathrm{K/cm}$. (c) and (d) Anisotropy function $a_n(\mathbf{n})$ for the excess interface free energy in different cross-sections.} \label{Fig:free_energy_anis}
\end{figure*}

\subsubsection{Formulation} \label{Sec:free_energy_form}

The free-energy anisotropy of the ice-water interface exhibits six-fold symmetry within the basal plane, including six preferred $\left<11\bar{2}0\right>$ directions ($a$-axis) and six prism $\left<1\bar{1}00\right>$ directions ($t$-axis), and has two cusps along the $\left<0001\right>$ directions ($c$-axis). We choose an anisotropy function $a_s(\mathbf{n})$ of the form:
\begin{equation}
a_s(\mathbf{n}) = a_s^0 \left(1 + \epsilon_{1} |\sin \theta| + \epsilon_{6} \sin^{6} \theta \cos 6 \varphi \right), \label{surf_energy_anis}
\end{equation}
where $\theta$ and $\varphi$ are the polar and azimuthal angles of the normal direction $\mathbf{n}$ in a standard spherical coordinate system [Fig.~\ref{Fig:free_energy_anis}(a)]. Here, the $z$ axis (with respect to which the polar angle $\theta$ is defined) is aligned with the $c$-axis, and the $x$-$y$ plane is parallel to the basal plane.
The second term inside the brackets introduces two cusps along the $\left<0001\right>$ directions, with $|\sin{\theta}|$ regularized by $\sqrt{\sin^2 \theta + d^2}$, where $d$ is a small regularization parameter set to $d = 0.01$ in simulations \cite{Wang2018Phase-fieldGrowth}. \added{The cusps in the free-energy anisotropy give rise to two small facets normal to the $c$-axis in the equilibrium shape (as shown later in this section), consistent with experimental observations that, under near-equilibrium growth conditions, ice crystals still exhibit two small facets normal to the $c$-axis~\cite{maruyama2011relation}.} The third term inside the brackets represents a spherical harmonic with hexagonal symmetry. 
The coefficients $\epsilon_1$ and $\epsilon_6$ are chosen based on the ice-water interfacial free energy estimated from various molecular dynamics models \cite{Davidchack2012IceInteractions}. Literature values give average free energies of $25.9~\mathrm{mJ/m^2}$ for the basal plane, and $28.0~\mathrm{mJ/m^2}$ and $29.1~\mathrm{mJ/m^2}$ for the interfaces perpendicular to the $\left<1\bar{1}00\right>$ and $\left<11\bar{2}0\right>$ directions, respectively. Accordingly, these values yield $\epsilon_1 = 0.1$, $\epsilon_6 = 0.02$, and $a_s^0 = 0.925$, with the reference value $a_s(\mathbf{n}) = 1$ corresponding to the prism direction. Using these coefficients, the plots of $a_s(\mathbf{n})$ as functions of the polar and azimuthal angles are shown in Fig.~\ref{Fig:free_energy_anis}(c)-(d). In 2D, the anisotropy function has the same form, with the azimuthal angle $\varphi = 0^{\circ}$ taken as constant.

The angles $\theta$ and $\varphi$ can be evaluated locally based on the phase field $\phi$. With the $c$-axis aligned with the $z$ direction and the $a$-axis with the $x$ direction in 3D, the expressions for $\theta$ and $\varphi$ are:
\begin{equation}
\theta = \sin^{-1} \left(\frac{\sqrt{\phi_x^2 + \phi_y^2}}{\sqrt{\phi_x^2 + \phi_y^2 + \phi_z^2}} \right),
\end{equation}
and
\begin{equation}
\varphi = \cos^{-1} \left(\frac{-\phi_x}{\sqrt{\phi_x^2 + \phi_y^2}} \right),
\end{equation}
where $\phi_i$ denotes the partial derivative $\partial \phi / \partial i$, with $i = x, y, z$. With the $c$-axis aligned with the $y$ direction and the $a$-axis with the $x$ direction in 2D, only the polar angle $\theta$ needs to be evaluated:
\begin{equation}
\theta = \sin^{-1} \left(\frac{\phi_x}{\sqrt{\phi_x^2 + \phi_y^2}} \right).
\end{equation}

The misorientation angles in the PF simulation are implemented through rotation matrix methods in 2D \cite{Tourret2015GrowthStudy} and 3D \cite{Tourret2017GrainStudy,song2023cell}, where $\phi_x$, $\phi_y$, and $\phi_z$ terms are modified based on the misorientation angles. In \cite{Yin2023HierarchicalTemplating,Ji2025ParityTemplating}, we implement one misorientation angle $\gamma_0$ in 2D and two misorientation angles $\alpha_0$ and $\gamma_0$ in 3D. By definition, $\gamma_0$ is the angle between the $a$-axis and the temperature gradient $G$ within the plane that contains both the $a$ and $c$ axes, and $\alpha_0$ is the angle between the $a$-axis and a reference direction (the projection of $G$) within the plane that contains both the $a$ and $t$ axes.

\subsubsection{Sharp cusp limit}

To evaluate whether the cusps in Eq.~\eqref{surf_energy_anis} introduce potential numerical issues in PF simulations, we examine the stability in the sharp cusp limit ($d \to 0$) by rewriting the PF equation at equilibrium in 2D:
\begin{eqnarray}
\tau(\mathbf{n}) \frac{\partial \phi}{\partial t} &=& W_0^2 \left( C_{xx} \partial_{xx} \phi + C_{yy} \partial_{yy} \phi + C_{xy} \partial_{xy} \phi \right) \nonumber \\
&& + \phi - \phi^3 + \Delta_e \lambda (1 - \phi^2)^2,
\end{eqnarray}
where $\Delta_e$ is the effective undercooling, and the coefficients are given by:
\begin{align}
C_{xx} &= a_s^2 + \cos^2\theta \left(a_s^{\prime\prime} a_s + {a_s^{\prime}}^2 \right), \\
C_{yy} &= a_s^2 + \sin^2\theta \left(a_s^{\prime\prime} a_s + {a_s^{\prime}}^2 \right), \\
C_{xy} &= 2 a_s^{\prime} a_s (\cos^2\theta - \sin^2\theta) \nonumber \\
&\quad - 2 \sin\theta \cos\theta \left(a_s^{\prime\prime} a_s + {a_s^{\prime}}^2 \right).
\end{align}
For simplicity, we consider $a_s(\mathbf{n})$ in the form:
\begin{equation}
a_s(\theta) = a_s^0 \left(1 + \epsilon_1 \sqrt{\sin^2\theta + d^2}\right).
\end{equation}
The derivatives of $a_s(\theta)$ are then:
\begin{equation}
a_s^{\prime}(\theta) = \frac{a_s^0 \epsilon_1 \sin\theta \cos\theta}{\sqrt{\sin^2\theta + d^2}},
\end{equation}
and
\begin{equation}
a_s^{\prime\prime}(\theta) = \frac{a_s^0 \epsilon_1 \left(d^2 \cos^2\theta - d^2 \sin^2\theta - \sin^4\theta\right)}{(\sin^2\theta + d^2)^{3/2}}.
\end{equation}

We analyze the sharp cusp limit along the $c$-axis $(\theta = 0^\circ)$, where $a_s=a_s^0 (1+\epsilon_1 d)$, $a_s^{\prime}=0$, $a_s^{\prime\prime}=a_s^0 \epsilon_1/d$, and $\tau(\mathbf{n})=\tau_c$. The most stringent constraint on the time step $\Delta t$, derived from a Von Neumann stability analysis, is given by:
\begin{equation}
\Delta t \leq \frac{\tau_c (\Delta x / W_0)^2}{2(C_{xx} + C_{yy})} \approx \frac{\tau_c (\Delta x / W_0)^2}{2 a_s^0} \frac{d}{\epsilon_1 + d(2 + \epsilon_1^2)},
\end{equation}
where second-order or higher-order terms in $d$ have been neglected. For $\Delta x / W_0 = 0.8$ and other coefficients given in Sec.~\ref{Sec:free_energy_form}, this constraint yields $\Delta t \leq 0.0288 \tau_c$. With $\tau_c$ determined by the kinetic anisotropy used in this paper, this constraint is less stringent than the threshold for numerical instability in the diffusion equation and can be satisfied in most simulations. Thus, the cusps in the free-energy anisotropy in Eq.~\eqref{surf_energy_anis} will not introduce numerical issues in PF simulations.

\subsubsection{Equilibrium shape}

\begin{figure}[htb!]
\includegraphics[scale=1]{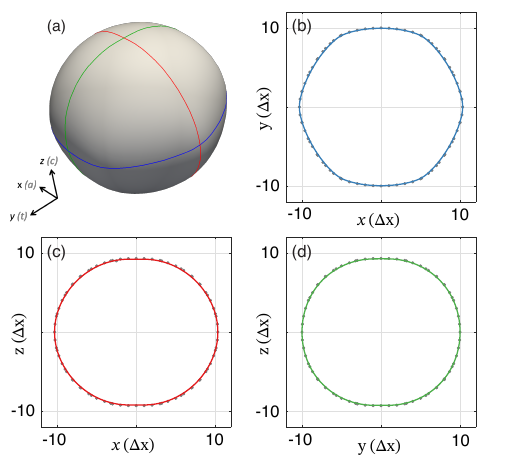}
\caption{(a) The numerically calculated equilibrium shape. (b)–(d) Contours of the equilibrium shape in different cross-sections (dots) compared with the 3D Wulff shape estimated using the $\boldsymbol{\xi}$-vector (lines).} \label{Fig:equilibrium_shape}
\end{figure}

From the free-energy anisotropy in Eq.~\eqref{surf_energy_anis}, we numerically determine the equilibrium shape of a solid with finite volume by incorporating the constraint $\int g(\phi) \, dV = C$ into a free-energy functional at $T = T_M$:
\begin{eqnarray}
F[\phi] = &&\int_{dV} \left[\frac{\sigma}{2} |\vec{\nabla} \phi|^{2} + f\left(\phi, T_M\right)\right] \nonumber \\
&&- \lambda_1 \left(\int_{dV} g(\phi) - C \right), \label{F_functional_equil}
\end{eqnarray}
where $\lambda_1$ is a dimensional Lagrange multiplier, and $C$ is a constant. The Lagrange multiplier ensures that the equilibrium shape neither grows nor shrinks. 
Substituting this free-energy functional into the dynamics of Eq.~\eqref{eq_phi} and using $g(\phi)$ from Eq.~\eqref{interpolation_g}, we obtain:
\begin{eqnarray}
\tau (\mathbf{n}) \frac{\partial \phi}{\partial t} = &&\vec{\nabla} \cdot \left[a_s(\mathbf{n})^{2} \vec{\nabla} \phi \right] + \tilde{\lambda}_1 (1 - \phi^2)^2 \nonumber \\
&&+ \sum_{m} \partial_{m} \left(|\vec{\nabla} \phi|^{2} a_s(\mathbf{n}) \frac{\partial a_s(\mathbf{n})}{\partial (\partial_{m} \phi)}\right), \label{equilibrium_shape}
\end{eqnarray}
where $\tilde{\lambda}_1$ is the dimensionless Lagrange multiplier. 
The relaxation time $\tau (\mathbf{n})$ has the same form for vanishing reciprocal kinetics and is given by $\tau (\mathbf{n}) = \tau_0 a_s(\mathbf{n})^2$. The value of $\tilde{\lambda}_1$ is determined numerically by substituting $\partial \phi / \partial t$ from Eq.~\eqref{equilibrium_shape} into the condition for volume conservation:
\begin{eqnarray}
\frac{d}{dt} \int g(\phi) \, dV = \int g'(\phi) \frac{\partial \phi}{\partial t} \, dV = 0. \label{Solve_Multiplier}
\end{eqnarray}
In numerical calculations of the equilibrium shape, we start from a regular sphere of radius $20 \Delta x$, with the grid spacing chosen as $\Delta x = 0.8 W_0$. We solve Eq.~\eqref{equilibrium_shape}, and iteratively update the Lagrange multiplier $\tilde{\lambda}_1$ during the numerical calculation. The equilibrium shape is obtained when the average change of $\phi$ at a lattice point becomes smaller than $10^{-12}$, as shown in Fig.~\ref{Fig:equilibrium_shape}(a).


To validate the calculated equilibrium shape, we compare it with the Wulff shape in 3D. The Wulff shape is constructed using the capillary vector $\boldsymbol{\xi}$, which sweeps the surface of the shape \cite{Hoffman1972AJunctions,Cahn1974ASurfaces} and is defined in spherical coordinates as:
\begin{equation}
\boldsymbol{\xi} = \Gamma(\mathbf{n}) \hat{r} + \frac{\partial \Gamma(\mathbf{n})}{\partial \theta} \hat{\theta} + \frac{1}{\sin{\theta}} \frac{\partial \Gamma(\mathbf{n})}{\partial \varphi} \hat{\varphi},
\end{equation}
where $\Gamma(\mathbf{n}) = \Gamma_0 a_s(\mathbf{n})$ is the orientation-dependent excess interface free energy, with $\Gamma_0$ being a scaling parameter. 
Since the size of the equilibrium shape depends on the initial seed, we adjust the value of $\Gamma_0$ and thus the Wulff shape to match the equilibrium shape obtained from simulations. As shown in Fig.~\ref{Fig:equilibrium_shape}, the numerically computed equilibrium shape agrees well with the Wulff shape. However, in both shapes, only two very small facets are found normal to the $c$-axis. Thus, the weakly anisotropic free energy alone does not seem sufficient to explain the large facets formed during ice-crystal growth as confirmed by our out-of-equilibrium growth simulations without kinetic anisotropy (cf. Fig. 1 in \cite{Ji2025ParityTemplating} and results in section V).

\subsection{Anisotropy of atomic attachment kinetics} \label{Sec:kinetic_anisotropy}

The atomic attachment kinetics refers to the interfacial dynamic processes that govern how atoms or molecules from the disordered liquid state attach to the crystal lattice \cite{Woodruff1973TheInterface,Libbrecht2017}. The interface kinetics is sufficiently fast in atomically rough directions, and the interface can be assumed in local thermodynamic equilibrium. In contrast, the growth of faceted interfaces is a much slower process, typically involving layer-by-layer growth mechanisms controlled by 2D nucleation or spiral growth around screw dislocations \cite{Woodruff1973TheInterface}. 
This interface kinetics can be described by the relationship between the interface velocity, $V_n$, and the kinetic undercooling, $\Delta T_k$. We can rewrite Eq.~\eqref{Gibbs}, such that the interface undercooling $\Delta T$ is measured with respect to the liquidus temperature as $\Delta T = T_M - |m| c_l - T_I$, where $T_I$ is the interface temperature. The total undercooling $\Delta T$ consists of contributions from both capillary and kinetic undercooling, i.e., $\Delta T = \Delta T_c + \Delta T_k$, where $\Delta T_c = \Gamma_{\mathrm{GT}} \mathcal{K}$ represents the capillary undercooling. The kinetic relationship is described by:
\begin{equation}
V_n = \mu_k^{\left<0001\right>} \Delta T_k, \label{V_n_linear}
\end{equation}
where $\mu_k^{\left<0001\right>}$ denotes the kinetic coefficient along the $\left<0001\right>$ faceted growth directions. 
The kinetic coefficient $\mu_k^{\left<0001\right>}$ is generally a nonlinear function of $\Delta T_k$ over a wide range of undercooling, reflecting underlying mechanisms such as the processes of layer-by-layer growth. However, for a limited range of undercooling spanning the narrow tip region of growing ice crystals, $\mu_k^{\left<0001\right>}$ can be reasonably approximated as a constant that restores a linear kinetic relationship.

In the quantitative PF model, we incorporate anisotropic interface kinetics by selecting a form of the kinetic coefficient $\mu_k(\mathbf{n})$ that interpolates between a vanishingly small kinetic undercooling for atomically rough interface growth in directions within the basal plane (corresponding to a negligible value of $\beta_k$) and a finite value of $\mu_k^{\left<0001\right>}$ along the $c$-axis. Both linear and nonlinear interface kinetics along the $c$-axis can be reproduced by the PF model. 

\subsubsection{Linear kinetics} \label{Sec:linear_kinetics}

\begin{figure}[H]
\includegraphics[scale=1]{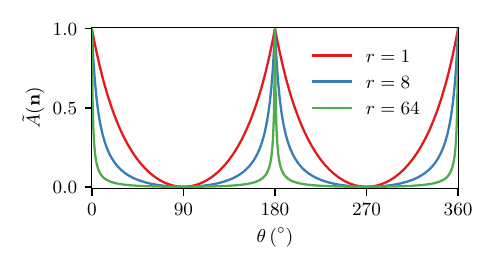}
\caption{The shape of the anisotropy function $\tilde{A}(\mathbf{n})$ for interface kinetics in the plane containing both the $a$ and $c$ axes. The three curves correspond to kinetic anisotropy slopes of $r = 1$, $r = 8$, and $r = 64$, respectively.} \label{Fig:kinetic_anis}
\end{figure}

To reproduce linear kinetic relationship on the basal plane, it is necessary to cancel the temperature dependence of $\beta_k$ in Eq.~\eqref{beta} along the $c$-axis. This is achieved by choosing a form of the relaxation time:
\begin{equation}
\tau(T, \mathbf{n}) = \tau_0 \left[ \tilde{c}_l^0(T) + A(\mathbf{n}) \right], \label{21_tau_T_n}
\end{equation}
where the anisotropy function $A(\mathbf{n})$ equals zero for directions perpendicular to the $c$-axis. Substituting Eq.~\eqref{21_tau_T_n} into Eq.~\eqref{beta}, we obtain:
\begin{equation}
\beta_k(\mathbf{n}) = \frac{a_1 a_2 W}{D} A(\mathbf{n}), \label{beta_linear}
\end{equation}
where $W$ is treated as a constant.
\added{Although the full angular shape of the kinetic anisotropy for the ice-water interface is not well established, we can choose an appropriate form of $A(\mathbf{n})$ that ensures finite kinetics along the $c$-axis while restoring atomically rough interfaces within the basal plane.}
A possible expression for $A(\mathbf{n})$ is given by:
\begin{equation}
A(\mathbf{n}) = A_0 \left[ \frac{1}{1 - r(1 - |\sin{\theta}|)/(1 + r)} - 1 \right], \label{A_linear}
\end{equation}
where $A_0$ is a scaling parameter, and $r$ is a constant that controls the variation of the interface kinetic coefficient with orientation. This choice also ensures that $1/\tau(T, \mathbf{n})$ has two cusps in the $\left<0001\right>$ directions, \added{reflecting a sharp change in the orientation‑dependent kinetics near the $c$-axis that is consistent with 2D nucleation or step‑flow growth on the basal plane \cite{mochizuki2023microscopic,murata2022step}}. Along the $c$-axis ($\theta = 0^\circ$), $A(\mathbf{n}) = A_0 r$, and the derivative $\partial A / \partial \theta$ is $[-A_0 r(1 + r)]$. In directions that are ontained within the basal plane ($\theta = 90^\circ$), $A(\mathbf{n}) = 0$ and $\beta_k$ vanishes.
To illustrate the behavior of the anisotropy function, we define a dimensionless quantity $\tilde{A}(\mathbf{n})$ as:
\begin{equation}
\tilde{A}(\mathbf{n}) \equiv \frac{A(\mathbf{n})}{A_0 r} = \frac{1 - |\sin{\theta}|}{1 + r |\sin{\theta}|}, \label{dimensionless_A}
\end{equation}
and plot its shapes for different values of $r$ in Fig.~\ref{Fig:kinetic_anis}.

With the choice of $A(\mathbf{n})$ in Eq.~\eqref{A_linear}, the reciprocal kinetic coefficient $\beta_k^{\left<0001\right>}$ on the basal plane depends on $W$, $A_0$, and $r$:
\begin{equation}
\beta_k^{\left<0001\right>} = \frac{a_1 a_2 W}{D} A_0 r. \label{beta_linear_0001}
\end{equation}
In the following PF simulations, we examine the convergence of the PF model as a function of $r$ and $W$. To change the shape of the kinetic anisotropy without affecting the magnitude of $\beta_k^{\left<0001\right>}$, $A_0$ should be adjusted such that the product $(A_0 r)$ remains constant. Similarly, to vary the interface thickness $W$ without altering the magnitude of $\beta_k^{\left<0001\right>}$, $A_0$ should be adjusted to keep the product $(W A_0)$ fixed.

\subsubsection{Nonlinear kinetics}

The kinetic coefficient $\mu_k^{\left<0001\right>}$ for the basal plane growth of ice crystals is generally a nonlinear function of $\Delta T_k$ over a wide range of undercooling. At relatively small kinetic undercooling, it can be expressed as:
\begin{equation}
\mu_k^{\left<0001\right>} = c_1 \exp\left(-\frac{c_2}{\Delta T_k}\right), \label{mu_k}
\end{equation}
where $c_1 = 7.3 \times 10^{-4}~\mathrm{m/(s \cdot K)}$ and $c_2 = 0.23~\mathrm{K}$ are constants fitted from experimental measurements \cite{Libbrecht2017}. Our goal is to reproduce this nonlinear kinetic relationship within the PF model. It is worth noting that the basal plane kinetics may be influenced by solutes \cite{Michaels1966ImpurityWater}, but this effect is expected to be negligible for dilute water-based solutions at small $\Delta T_k$ considered in this paper.

As discussed later in Sec.~\ref{Sec:basal_pane}, by substituting $c_l$ with $c_{\infty} e^u$, the kinetic undercooling at a faceted interface can be expressed as:
\begin{equation}
\Delta T_k = T_M - T - |m| c_l \approx -\Delta \tilde{c} \Delta T_0,
\end{equation}
where
\begin{equation}
\Delta \tilde{c} \equiv \frac{2 \tilde{c}}{1 - \phi + \delta} - \tilde{c}_l^0(T) \label{delta_c}
\end{equation}
represents the dimensionless deviation of the concentration from the equilibrium liquidus concentration at a temperature $T$.

To model the nonlinear kinetics on the basal plane, the kinetic coefficient in Eq.~\eqref{beta_linear} should also depend on $\Delta \tilde{c}$. For this purpose, we retain Eqs.~\eqref{21_tau_T_n}–\eqref{A_linear}, but the parameter $r$ in Eq.~\eqref{A_linear} is no longer constant. Instead, it takes the form:
\begin{equation}
r = r_0 \exp\left(-\frac{g}{\Delta \tilde{c}}\right), \label{21_r_exp}
\end{equation}
where $r_0 = f / A_0$, $f = D / (a_1 a_2 c_1 W \Delta T_0)$, and $g = c_2 / \Delta T_0$. 
Substituting Eq.~\eqref{21_r_exp} into Eq.~\eqref{A_linear}, we obtain the following form of anisotropy function for nonlinear kinetics:
\begin{equation}
A(\mathbf{n},\Delta \tilde{c})=A_0 \left[ \frac{1}{1-(1-|\sin{\theta}|)/\left(1+r_0^{-1} e^{{g}/{\Delta \tilde{c}}}\right)}-1 \right]. \label{A_nonlinear}
\end{equation}
The scaled anisotropy function remains the same as in Eq.~\eqref{dimensionless_A}, with $r$ now replaced by Eq.~\eqref{21_r_exp}. Consequently, its shape remains as shown in Fig.~\ref{Fig:kinetic_anis} for a fixed $\Delta \tilde{c}$. In PF simulations, $\Delta \tilde{c}$ is calculated using Eq.~\eqref{delta_c} at each time step, which provides feedback to determine the interface kinetics. Since $\Delta \tilde{c}$ appears in the denominator of the exponential term in Eq.~\eqref{A_nonlinear}, a cutoff is imposed during numerical simulations to ensure that its absolute value remains larger than a small threshold, set to $10^{-4}$.

With the choice of $A(\mathbf{n},\Delta \tilde{c})$ in Eq.~\eqref{A_nonlinear}, the reciprocal kinetic coefficient on the basal plane is given by:
\begin{equation}
\beta_k^{\left<0001\right>}(\Delta \tilde{c}) = \frac{1}{c_1 \Delta T_0} \exp\left(-\frac{g}{\Delta \tilde{c}}\right). \label{beta_nonlinear}
\end{equation}
Here, $\beta_k^{\left<0001\right>}$ is independent of $W$ and $A_0$. The scaling parameter $A_0$ only affects the rate of kinetic coefficient variation in directions close to the $c$-axis through the value of $r_0$. Meanwhile, in the directions parallel to the basal plane, $\beta_k$ still vanishes with the anisotropy function in given Eq.~\eqref{A_nonlinear}.

\section{Numerical results} \label{sec:numerical_results}

In this section, we use the presented PF model to simulate the directional solidification of water-based solutions and perform additional analyses. Considering the anisotropic properties of the interface, both $W$ and $\tau$ in the PF model are orientation-dependent: the interface thickness is expressed as $W(\mathbf{n}) = W_0 a_s(\mathbf{n})$, where the anisotropy function $a_s(\mathbf{n})$ is introduced in Sec.~\ref{Sec:free_energy}; similarly, as discussed in Sec.~\ref{Sec:kinetic_anisotropy}, the relaxation time is given by $\tau(T, \mathbf{n}) = \tau_0 \left[\tilde{c}_l^0(T) + A(\mathbf{n})\right]$. The final evolution equations with both anisotropies are:
\begin{eqnarray}
&&\left[ \tilde{c}_l^0(T)+ A(\mathbf{n}) \right] a_s(\mathbf{n})^{2} \frac{\partial \phi}{\partial t} = \vec{\nabla} \cdot\left[a_s(\mathbf{n})^{2} \vec{\nabla} \phi\right] \nonumber \\
&&\qquad+\sum_{m}\left[\partial_{m}\left(|\vec{\nabla} \phi|^{2} a_s(\mathbf{n}) \frac{\partial a_s(\mathbf{n})}{\partial\left(\partial_{m} \phi\right)}\right)\right] \label{dmodel_phi} \\
&&\qquad+\phi-\phi^{3}-\lambda (1-\phi^2)^2 \left[\frac{2 \tilde{c}}{1-\phi+\delta}-\tilde{c}_l^0(T)\right], \nonumber
\end{eqnarray}
\begin{eqnarray}
\frac{\partial \tilde{c}}{\partial t}=&& \widetilde{D} \vec{\nabla} \cdot \left( \tilde{c} \vec{\nabla} \ln \frac{\tilde{c}}{1-\phi+\delta} \right) \nonumber\\
&&-\frac{1}{\sqrt{2}} \vec{\nabla} \cdot \left[ \tilde{c} \, \partial_t \ln (1-\phi+\delta) \frac{\vec{\nabla} \phi}{|\vec{\nabla} \phi|} \right], \label{dmodel_c}
\end{eqnarray}
where $\tilde{c} = c / c_{\infty}$ is the dimensionless solute concentration, and time and length are scaled by $\tau_0$ and $W_0$, respectively. The dimensionless coefficients include $\lambda = a_1 W_0 / d_0$ and $\widetilde{D} = D \tau_0 / W_0^2 = a_1 a_2 W_0 / d_0$, where $a_1 = 5 \sqrt{2} / 8$ and $a_2 = 47 / 75$ are the same numerical constants as in Refs.~\cite{karma1996phase,Karma1998QuantitativeDimensions,Echebarria2004QuantitativeSolidification}. \added{The equilibrium liquidus $c_l^0(T)$ enters the PF model through $\tilde{c}_l^0(T) \equiv c_l^0(T)/c_\infty$ in Eq.~\eqref{dmodel_phi}. The present PF model accommodates either a linear or nonlinear liquidus by appropriately choosing the input form of $\tilde{c}_l^0(T)$. For a straight liquidus and directional solidification under the frozen temperature approximation, $\tilde{c}_l^0(T)$ is replaced by Eq.~\eqref{directional_c}.}

We consider the directional solidification of dilute sucrose- and trehalose-water solutions, which represent binary aqueous mixtures with small solutes obeying Fickian diffusion. The freeze-cast materials produced from these simple systems exhibit characteristic hierarchical lamellar structures that closely resemble those observed in more complex systems \cite{Yin2023HierarchicalTemplating}, indicating that the structural formation is mainly diffusion-controlled. Sucrose and trehalose, having the same molar mass, are treated as having the same properties in PF simulations. The physical parameters of these solutions, along with processing conditions, are summarized in Table~\ref{tab:table1}. The chosen pulling velocity $V_p = 15~\mathrm{\mu m/s}$ and temperature gradient $G = 12~\mathrm{K/cm}$ represent typical processing conditions during ice-templating experiments \cite{Yin2023HierarchicalTemplating}, which are used in all simulations in this paper. 
For the kinetic coefficient $\mu_k^{\left<0001\right>}$ on the basal plane, unless otherwise specified, we choose a value of $41.1\,\mathrm{\mu m/s/K}$, which lies within the range $\mu_k^{\mathrm{min}} \leq \mu_k^{\left<0001\right>} \leq \mu_k^{\mathrm{max}}$, where spontaneous symmetry breaking that leads to the formation of partially faceted ice lamellae occurs \cite{Ji2025ParityTemplating}.
The PF simulations capture the dynamics of the solidification front over a temperature range where the solute concentration remains dilute and the liquidus slope is quantitatively predicted by the analytical Clausius-Clapeyron relation \cite{Karma1993}, with details given in Appendix \ref{Sec:phase_diagram}. Although this temperature range is above the vitrification temperature, it includes a sufficiently large region of the solidification front to capture key morphological instabilities that shape the hierarchical structures of ice crystals and the resulting templated materials.

\begin{table}[hbt!]
\caption{\label{tab:table1}%
Materials and processing parameters for directional solidification of the trehalose-water solution.
}
\begin{ruledtabular}
\begin{tabular}{llll}
\textrm{Symbol} & \textrm{Parameter} & \textrm{Value} & \textrm{Unit} \\
\colrule
$c_{\infty}$ & Nominal concentration & 3 & $\mathrm{wt.\%}$ \\
$m$ & Liquidus slope & -0.0543 & $\mathrm{K / wt.\%}$ \\
$D$ & Diffusion coefficient & 140.7 & $\mathrm{\mu m^2 / s}$ \\
$\Gamma_{\mathrm{GT}}$ & Gibbs-Thompson coefficient & $1.64 \times 10^{-8}$ & $\mathrm{K\cdot m}$ \\
$d_0$ & Capillary length & 0.1 & $\mathrm{\mu m}$ \\
$l_D$ & Diffusion length & 9.38 & $\mathrm{\mu m}$ \\
$l_T$ & Thermal length & 135.6 & $\mathrm{\mu m}$ \\
$V_p$ & Pulling speed & 15 & $\mathrm{\mu m/s}$ \\
$G$ & Temperature gradient & 12 & $\mathrm{K/cm}$ \\
\end{tabular}
\end{ruledtabular}
\end{table}

The evolution equations \eqref{dmodel_phi}–\eqref{dmodel_c} are solved on a square lattice in 2D and a cubic lattice in 3D using the finite-difference method with grid spacing $\Delta x$ and the explicit Euler method with a time step $\Delta t$. A grid spacing of $\Delta x = 0.8 \, W_0$ is used in all simulations, and the time step is chosen to remain below the threshold of numerical instability for the diffusion equation, i.e., $\Delta t/\tau_0$ is smaller than ${(\Delta x/W_0)^2}/{4 \widetilde{D}}$ in 2D and ${(\Delta x/W_0)^2}/{6 \widetilde{D}}$ in 3D. The PF model is implemented for massively parallel computing on Nvidia Tesla V100 GPUs using the CUDA programming language. The time loop is achieved by swapping the pointer addresses of arrays containing the $\phi$ and $\tilde{c}$ fields at the current and next time steps. Additional details about the spatial discretization of the evolution equations are provided in Appendix~\ref{Sec:Dis_evo_eqs}. 

Initially, the planar solid-liquid interface perpendicular to $G$ is positioned at its steady-state location where $T = T_0$. The $\phi$ field at position $x$ is initialized as:
\begin{equation}
\phi(z) = \tanh{\left(\frac{x_0 - x}{\sqrt{2}}\right)},
\end{equation}
where $x_0$ is the initial position of the interface, in units of $W$. The $\tilde{c}$ field is initialized as:
\begin{equation}
\tilde{c}(x) = \frac{1 - \phi(x) + \delta}{2},
\end{equation}
where $\delta = 10^{-6}$ is a small constant used in all simulations. This initial condition ensures a smooth variation of the concentration field from $c_{\infty}$ in the liquid phase to a small value ($\delta/2$) in the solid phase. An initial perturbation $\eta_0 F_\eta(\vec{r})$ is applied perpendicular to the interface (along the $x$-axis), where $\eta_0 = 0.5 \Delta x$ is the noise amplitude, and $F_\eta(\vec{r})$ is a random function of $\vec{r}$ along the horizontal interface. The values of $F_\eta(\vec{r})$ are are generated randomly from a uniform distribution between $[-0.5,\,0.5]$.
2D PF simulations are performed in a plane containing both the $a$ and $c$ axes, where periodic boundary conditions are applied in the horizontal directions. Although 2D simulations neglect secondary dynamics within an ice lamella, they effectively capture the primary lamellar dynamics observed in 3D. 
In 3D simulations, periodic boundary conditions are applied in both the transverse and $\left<1\bar{1}00\right>$ directions. 
The most advanced solid-liquid interface (ice tip) is maintained at a fixed $x$ location by pulling back the entire simulation domain and truncating the excess portion of the $\phi$ and $\tilde{c}$ fields at the rear $-x$ boundary. Here, no-flux boundary conditions are implemented at the $\pm x$ boundaries.

\subsection{Effects of anisotropies}

We perform 3D PF simulations to investigate the effects of both anisotropic interfacial properties on microstructural pattern formation. For small solute particles such as sucrose and trehalose molecules, Brownian diffusion dominates, and the instabilities at the growth front of ice crystals resemble the classical Mullins-Sekerka instability observed in alloy solidification. As the planar interface breaks down, these instabilities lead to protuberances with small wavelengths \cite{Yin2023HierarchicalTemplating,Ji2025ParityTemplating}. Subsequently, the development of these initial instabilities into larger-scale microstructures is significantly influenced by the anisotropic properties of the interface.

\begin{figure}[hbt!]
\includegraphics[scale=0.9]{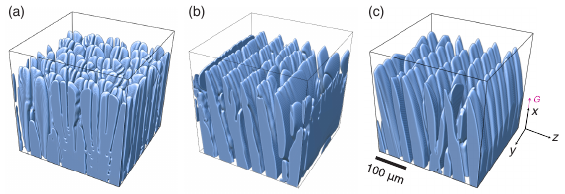}
\caption{Ice crystals in 3D PF simulations of the directional solidification of a 3 wt.\% aqueous sucrose solution under growth conditions of $V_p = 15~\mathrm{\mu m/s}$ and $G = 12~\mathrm{K/cm}$: (a) with only free-energy anisotropy, (b) with only kinetic anisotropy, and (c) with both anisotropies. In all simulations, the $\left<11\bar{2}0\right>$ preferred growth direction is aligned with the temperature gradient $G$, which is parallel to the $x$-axis of the rectangular coordinate system, while the $\left<0001\right>$ direction is parallel to the $z$-axis. \label{fig:3D_PF_anisotropy}}
\end{figure}

As shown in the PF simulation with only weak anisotropy in interface free energy [Fig.~\ref{fig:3D_PF_anisotropy}(a)], i.e., setting $\beta_k$ to vanish in all directions, the initial instabilities evolve into columnar cells without prominent facets. These cells grow in a non-steady state, exhibiting continuous tip splitting. In this case, the parity symmetry of the solidification front remains unbroken. 
In contrast, when only strong anisotropy in interface kinetics is included and the free energy is made isotropic, as shown in Fig.~\ref{fig:3D_PF_anisotropy}(b), the initial instabilities evolve into lamellar structures, and parity symmetry is spontaneously broken. These lamellar structures exhibit a faceted surface perpendicular to one of the $\left<0001\right>$ directions and a rough surface on the opposite side. Secondary instabilities occur on the rough surface of the ice lamellae. However, due to the absence of interface free-energy anisotropy, the solidification front does not select steady-state cellular tips within the ice lamellae. Consequently, the unilateral surface features on the rough side of the lamellar structure are highly disordered.
Lastly, in the PF simulation incorporating both anisotropic interfacial properties [Fig.~\ref{fig:3D_PF_anisotropy}(c), and Fig.~1(d) in \cite{Ji2025ParityTemplating}], the initial instabilities evolve into partially faceted lamellar structures. Steady-state ice tips are selected within the primary ice lamellae, leading to unilateral surface features on the rough side of the freeze-casted material, such as secondary ridges with spacings predicted by PF simulations that agree quantitatively with experimental observations \cite{Yin2023HierarchicalTemplating}.
Additionally, the less frequent tip elimination instabilities give rise to jellyfish-like substructures on the rough side \cite{Yin2023HierarchicalTemplating}.

The separate effects of interface free-energy and kinetic anisotropies are thus evident. The highly anisotropic interface kinetics is crucial for spontaneous symmetry breaking, leading to the formation of partially faceted ice lamellae. While weakly anisotropic interface free energy cannot independently induce the formation of partially faceted structures, it controls the solidification front and dictates the formation of unilateral surface features on the rough side of the lamellae. The role of the $\epsilon_6$ term in Eq.~\eqref{surf_energy_anis} is analogous to the effect of the free-energy anisotropy in alloy solidification. While $\epsilon_6$ selects ice tips that are contained within a thin lamellar structure, the latter selects the rounded tips of 3D cellular and dendritic structures in alloy solidification. Additionally, the cusps in the free-energy anisotropy introduced by the $\epsilon_1$ term in Eq.~\eqref{surf_energy_anis} enhance the stability of the facet and the lamellar structure. Without these cusps, the elongation of ice lamellae in the $\left<1\bar{1}00\right>$ direction becomes shorter and the lamellar structure tends to be more disordered.
In summary, both interface free-energy anisotropy and kinetic anisotropy have distinct yet important effects, and their interplay collectively leads to the hierarchical structure formation observed in experiments. Therefore, the accurate implementation of both anisotropic interfacial properties is essential for quantitative PF modeling of ice templating. All simulations in the following have both anisotropies incorporated.

\added{As the PF simulations in Fig.~\ref{fig:3D_PF_anisotropy} demonstrate, highly anisotropic interface kinetics have predominant effects on the formation of the partially faceted lamellar structure, whereas the weakly anisotropic free-energy anisotropy plays a secondary role. This result also suggests that the microscopic solvability theory \cite{Langer1980InstabilitiesGrowth,brener1991pattern}, which is generally developed for weakly anisotropic interfaces, cannot be directly applied here. A solvability theory for this kinetics-driven, highly faceted system (and its comparison with PF simulations) warrants further investigation in future studies.}

\subsection{Faceted ice-crystal growth}

\begin{figure}[hbt!]
\includegraphics[scale=1]{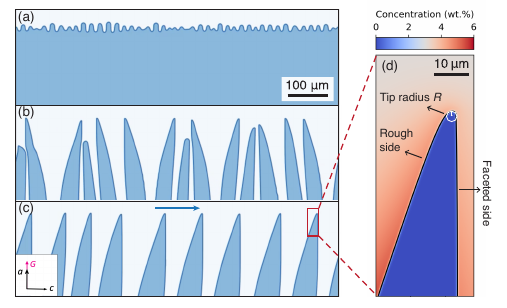}
\caption{Ice crystals at time $t=$ 25 s (a), 66 s (b), and 400 s (c), captured in a 2D PF simulation of the directional solidification of a 3 wt.\% aqueous sucrose solution under growth conditions of $V_p = 15~\mathrm{\mu m/s}$ and $G = 12~\mathrm{K/cm}$. The solid-liquid interface is initially planar, in a steady-state at rest, and located at the liquidus temperature. (d) A zoomed-in image of the tip region, where the colormap represents the solute concentration. \added{The blue arrow indicates the drifting direction.}\label{fig:figure_2D}}
\end{figure}

We perform 2D PF simulations in the $x$-$z$ plane that contains both the $a$ and $c$ axes to investigate the morphology and growth dynamics of ice crystals. In a spatially extended system, as shown in Fig.~\ref{fig:figure_2D}, initial morphological instabilities develop into ice lamellae consisting of two branches with opposite drifting directions (parallel to the $c$-axis) that compete with each other. Since these two branches are equivalent, neither branch has a significant advantage during the competition process. In a finite simulation domain, one branch eventually survives after a stochastic process, which takes considerably longer than the selection process for two branches with a misorientation angle $\gamma_0 > 0^{\circ}$ \cite{Ji2025ParityTemplating} that follows the classic Walton and Chalmers minimum undercooling criterion \cite{walton1959origin}.
In the steady-state array that drifts laterally along the $c$-axis, the ice lamellae exhibit partially faceted structures with sharp tips of radius $R$, as shown in Fig.~\ref{fig:figure_2D}(d). 

\subsubsection{Convergence of the drifting velocity} \label{sec:converge_Vd}

To assess the convergence of the PF model, we first measure the drifting velocities of steady-state ice lamellae for various interface thicknesses $W_0$. Both linear and nonlinear kinetics for basal plane growth are considered.
For linear kinetics, since $\beta_k^{\left<0001\right>}$ on the basal plane depends on $W_0$, as described in Eq.~\eqref{beta_linear_0001}, we adjust $A_0$ for each $W_0$ to maintain a fixed value of the kinetic coefficient for basal plane growth ($\mu_k^{\left<0001\right>} = 41.1 \, \mathrm{\mu m/s/K}$). For nonlinear kinetics, however, the anisotropy function does not require adjustment for different $W_0$ values, as $\beta_k^{\left<0001\right>}$ is independent of $W_0$ shown in Eq.~\eqref{beta_nonlinear}. Additionally, the grid size is varied with $W_0$ to ensure that the simulation domain maintains consistent dimensions across all simulations. For these simulations, we choose a lamellar spacing of $\lambda=60~\mathrm{\mu m}$ within the stable range of $\lambda$ and focus on investigating the dynamics of a single ice lamella in 2D PF simulations. To ensure only one lamella is contained within the simulation domain, a sinusoidal perturbation of small amplitude is added to the initial planar interface, with a wavelength equal to the primary spacing $\lambda$ and its maximum located at the boundary. After the transient period, a single steady-state lamella is selected, drifting laterally at a constant velocity $V_d$ within the simulation domain.

\begin{figure}[hbt!]
\includegraphics[scale=1]{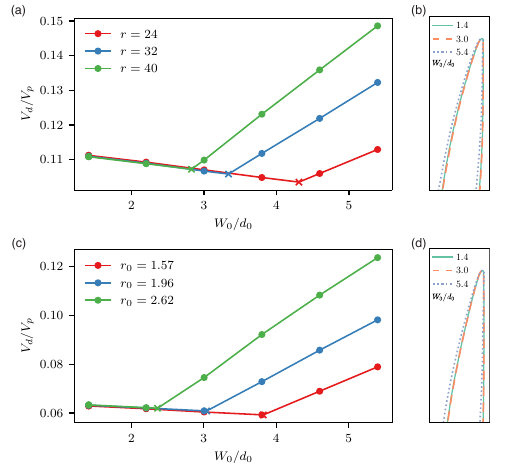}
\caption{(a) Convergence of the drifting velocity $V_d$ in 2D simulations with linear kinetics, where different lines represent various values of $r$. (b) Comparison of ice lamellae in simulations with different interface thicknesses $W_0$ for linear kinetics at $r = 32$. (c) Convergence of $V_d$ in simulations with nonlinear kinetics, where different lines represent various values of $r_0$. 
(d) Comparison of ice lamellae in simulations with different $W_0$ for nonlinear kinetics at $r_0 = 1.96$. 
The lamellar spacing is fixed at $60~\mathrm{\mu m}$ in all simulations.}  \label{Fig:convergence_v}
\end{figure}

\begin{table}[hbt!]
\caption{\label{tab:table2}
The critical interface thickness $W_c/d_0$ where the drifting velocity of an ice lamella converges sharply.
}
\begin{ruledtabular}
\begin{tabular}{cc}
\textrm{$r$ (linear kinetics)}&
\textrm{$W_c/d_0$}\\
\colrule
24 & 4.31\\
32 & 3.34\\
40 & 2.83\\
\hline
\textrm{$r_0$ (nonlinear kinetics)}&
\textrm{$W_c/d_0$}\\
\hline
1.57 & 3.82\\
1.96 & 3.34\\
2.26 & 2.36\\
\end{tabular}
\end{ruledtabular}
\end{table}

The plots of $V_d/V_p$ as functions of $W_0/d_0$ in Fig.~\ref{Fig:convergence_v} show the convergence of PF simulations for both linear and nonlinear kinetics, where different shapes of the kinetic anisotropy are also compared. In both cases, $V_d$ converges sharply at a critical interface thickness, $W_c$, beyond which the variation of $V_d$ becomes negligible as $W_0/d_0$ decreases. The morphologies of the ice lamellae also converge well for $W_0<W_c$, as illustrated in panels (b) and (d) of Fig.~\ref{Fig:convergence_v}. Since the variations of $V_d/V_p$ as functions of $W_0/d_0$ can be approximated by two linear functions for small and large $W_0/d_0$, $W_c$ is determined by fitting these functions and identifying their intersection. The measurements of $W_c$ are summarized in Table~\ref{tab:table2}. The results indicate that $W_c$ depends on the slope of the variation of the kinetic anisotropy with orientation near the $c$-axis, i.e., $r$ in the anisotropy function for linear kinetics and $r_0$ for nonlinear kinetics. For smaller slopes, $V_d$ converges more readily at larger values of $W_c$. Once convergence is achieved, the lamellae drifting velocity $V_d$ becomes independent of $r$ or $r_0$, indicating that the rate of variation of the kinetic anisotropy with orientation does not affect the drifting dynamics of partially faceted ice lamellae that are primarily controlled by basal plane kinetics.

\begin{figure}[hbt!]
\includegraphics[scale=1]{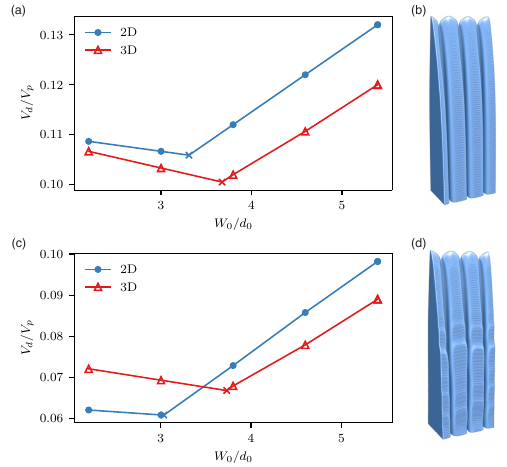}
\caption{(a) Convergence of the drifting velocity $V_d$ in 2D and 3D PF simulations for linear kinetics at $r = 32$. (b) 3D ice crystal corresponding to the data point $W_0 / d_0 = 3$ in (a). (c) Convergence of $V_d$ in 2D and 3D PF simulations for nonlinear kinetics at $r_0 = 1.96$. (d) 3D ice crystal corresponding to the data point $W_0 / d_0 = 3$ in (c). 
The lamellar spacing is fixed at $60~\mathrm{\mu m}$ in all simulations.}  \label{fig:convergence_3D}
\end{figure}

We also perform convergence tests in 3D PF simulations, initializing them with steady-state 2D solutions elongated along the perpendicular $\left<1\bar{1}00\right>$ direction with a lamellar spacing of $\lambda=60~\mathrm{\mu m}$. As shown in Fig.~\ref{fig:convergence_3D}, we choose a simulation domain of size 80~$\mathrm{\mu m}$ in the elongation direction, where the 3D ice lamellae contain three tips in steady state. With either linear or nonlinear kinetics on the basal plane, the morphologies of the partially faceted ice lamellae are similar, as shown in panels (b) and (d) of Fig.~\ref{fig:convergence_3D}. However, the rough side exhibits more pronounced undulations in the case of nonlinear kinetics. The drifting velocity, $V_d$, for 3D ice lamellae also converges sharply at a critical interface thickness $W_c$, with $W_c$ slightly larger than in the corresponding 2D simulations. This difference is likely due to the presence of additional orientations on the rough side of the 3D ice lamellae, which connect smoothly to the faceted side at the solidification front.
These results suggest that 2D simulations provide a reasonable approximation of the dynamics of ice lamellae. Thus, we primarily rely on 2D simulations for investigating the dynamics of primary lamellar structure that are more computationally efficient.

\subsubsection{Convergence of the tip radius and undercooling}

\begin{figure}[hbt!]
\includegraphics[scale=1]{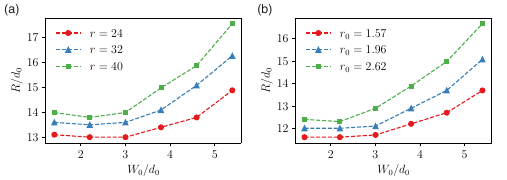}
\caption{Convergence of the tip radius $R$ in 2D simulations for (a) linear kinetics at different values of $r$, and (b) nonlinear kinetics at different values of $r_0$.} \label{Fig:convergence_r1}
\end{figure}

We perform 2D PF simulations to examine the effects of interface thickness and the shapes of kinetic anisotropy on the convergence of the ice tip radius $R$. The tip that connects smoothly the faceted side and the rough side of ice lamellae is generally much smaller than the scale of the lamellar structure, as shown in Fig.~\ref{fig:figure_2D}(d). 
The measurement of $R$ in PF simulations involves three steps: First, the contour of $\phi = 0$ is identified using the output of the $\phi$-field. Next, the most advanced interface position is located through interpolation to determine the tip position $(x_\mathrm{tip}, y_\mathrm{tip})$. Finally, the tip radius $R$ is extracted by fitting the rough side of the interface to a parabola, $y = y_\mathrm{tip} - (x - x_\mathrm{tip})^2 / (2R)$. 
As shown in Fig.~\ref{Fig:convergence_r1}, the tip radius converges more slowly at smaller $W$ values compared to the convergence of the drifting velocity $V_d$ for both linear and nonlinear kinetics. When the interface thickness is comparable to or larger than the tip radius, the simulation cannot accurately resolve the tip region. Since the tip radius is small, i.e., a few microns, a small $W_0/d_0 \approx 3$ is generally required to obtain well-converged results. It is worth noting that, unlike $V_d$, the converged $R$ is influenced by $r$ or $r_0$: a steeper slope results in a larger tip radius. This indicates that the non-faceted interface is sensitive to the form of the kinetic anisotropy, \added{and thus the convergence of the tip radius as a function of $r$ or $r_0$ will also be investigated next.}

While the form of the kinetic anisotropy is not precisely known, its slope near the $c$-axis can be approximately estimated using Eq.\eqref{beta_linear} and the kinetic coefficient $\mu_k$ in an atomically rough direction. Experimental data on the kinetics of atomically rough directions of ice-water interfaces are more difficult to obtain than data for basal plane growth \cite{Libbrecht2017}. Experimental and molecular dynamics studies of ice-water growth kinetics are typically conducted at large undercooling \cite{pruppacher1967interpretation,rozmanov2012isoconfigurational,buttersack2016critical,montero2023kinetics}, i.e., $\Delta T_k > 1$ K. By extrapolating these data, we estimate $\mu_k$ for atomically rough interfaces to be in the range of 2000 to 6000~$\mathrm{\mu m/s/K}$ for a small $\Delta T_k$, typically around $0.05~\mathrm{K}$, on the facet of ice lamellae.
Using this estimated range of $\mu_k$ as the kinetic coefficient for the ice-water interface at $\theta = \pi / 4$ (between the $a$-axis and $c$-axis) and assuming $\mu_k = 41.1~\mathrm{\mu m/s/K}$ on the basal plane, we solve for $r$ using Eq.\eqref{beta_linear}. This yields $r$ in a broad range from 8 to 60. A large value of $r$ in this range would require a very small $W_c$ for the convergence of lamellae drifting, as shown in Fig.\ref{Fig:convergence_v}, which poses significant challenges for numerical simulations.

\begin{figure}[hbt!]
\includegraphics[scale=1]{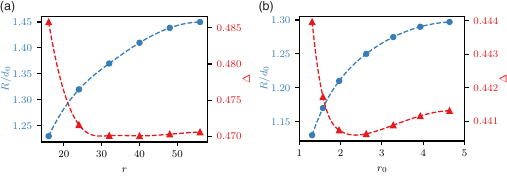}
\caption{The tip radius $R$ and undercooling $\Delta$ as functions of (a) $r$ for linear kinetics and (b) $r_0$ for nonlinear kinetics on the basal plane. A small interface thickness $W_0 / d_0 = 1.4$ is used in all simulations to ensure convergence of the drifting velocity.} \label{Fig:convergence_r2}
\end{figure}

To select a slope value within the estimated range for PF simulations, we analyze how ice-crystal growth converges with varying slopes $r$ and $r_0$ for linear and nonlinear kinetics, respectively. Since kinetics on the basal plane is unaffected by slopes for $W < W_c$, as shown in Sec.~\ref{sec:converge_Vd}, we measure the tip radius $R$ and dimensionless undercooling $\Delta$ for various values of $r$ and $r_0$ under the same basal plane kinetics. Here $\Delta$ is defined as
\begin{equation}
\Delta=\frac{T_M-T_\mathrm{tip}}{\Delta T_0}-1,
\end{equation}
where $T_\mathrm{tip}$ is the temperature at the most advanced point of the interface in the growth direction imposed by temperature gradient. The basal plane kinetics and $W_0$ are kept constants by adjusting $A_0$ in the anisotropy functions for both linear and nonlinear kinetics. A small interface thickness ($W_0/d_0 = 1.4$) is chosen to ensure converged $V_d$ across all tested slopes.
As shown in Fig.~\ref{Fig:convergence_r2}, the tip radius increases with the slope until an equilibrium value is reached, while the tip undercooling converges more rapidly than the tip radius at relatively small slopes. This demonstrates that we can safely select a slope that is small enough to be resolved by the model yet large enough to achieve converged tip radius and undercooling. Since the convergence of $V_d$ depends on the slope, and smaller $W_0$ is needed for steeper slopes, choosing a relatively small slope ensures computational efficiency. For the current set of parameters, we choose $r = 32$ and $r_0 = 1.96$ for linear and nonlinear kinetics, respectively, which satisfy these conditions.

These constraints on interface thickness and the slope of kinetic anisotropy may be stringent for PF simulations in 3D. In PF studies of ice templating \cite{Yin2023HierarchicalTemplating,Ji2025ParityTemplating}, spatially extended simulations in 3D are necessary to investigate microstructural pattern formation at experimentally relevant length and time scales. In such cases, satisfying strict convergence criteria is computationally expensive and may not be strictly required. A relatively large interface thickness of $W_0 / d_0 \approx 6$ is used to relax these constraints. In spatially extended 2D simulations, we select $W_0 / d_0 < 3$ for the same $r$ and $r_0$ values, ensuring convergence of $V_d$, $R$, and $\Delta$. By choosing these modeling parameters, we ensure that the PF simulations remain quantitative while maintaining computational efficiency.

\subsection{Basal plane kinetics} \label{Sec:basal_pane}

The dynamics of the partially faceted lamellar structure is primarily controlled by the interface kinetics on the basal plane. Here, we provide a detailed investigation of the basal plane kinetics by comparing the measured and imposed kinetic relationship in PF simulations. On the facet, the capillary effect can be neglected, and the interface temperature satisfies:
\begin{equation}
T = T_M - |m| c_l - \Delta T_k, \label{sharp_interface_Tk}
\end{equation}
where $T$ is known at a fixed position under the frozen temperature approximation. Thus, only $c_l$ is required to solve Eq.~\eqref{sharp_interface_Tk} and estimate $\Delta T_k$ on the facet. However, since the solute concentration varies rapidly across a diffuse interface in the complete-partitioning limit, it is difficult to directly measure the solute concentration $c_l$ on the liquid side of the interface in PF simulations.

Alternatively, we estimate $\Delta T_k$ using the chemical potential $\mu = \partial f_{AB} / \partial c$, which remains approximately constant at the interface. According to the definition of $f_{AB}$ in Eq.~\eqref{f_AB}, the chemical potential at equilibrium is given by:
\begin{equation}
\mu_E(T) = \frac{R_0 T_M}{v_0} \ln{c^0_l(T)}.
\end{equation}
Out of equilibrium, the chemical potential is expressed as:
\begin{equation}
\mu(\phi, c) = \frac{R_0 T_M}{v_0} \ln{c} + \epsilon(\phi).
\end{equation}
In addition, we define a dimensionless quantity $\bar{u}$, the measure of the departure of the chemical potential from its equilibrium value for a flat interface at a given temperature:
\begin{equation}
\bar{u} \equiv \frac{v_0}{R_0 T_M} \left( \mu - \mu_E \right) = \ln{\frac{2c / c^0_l(T)}{1 - g(\phi) + \delta}}. \label{chem_devi_ubar}
\end{equation}
The quantity $\bar{u}$ is used to derive an expression for the kinetic undercooling along spatially extended facets that form on one side of growing ice lamellae,
and there is a relation
\begin{equation}
e^{(u-\bar{u})}=c^0_l(T)/c_{\infty} \label{u_u_bar}
\end{equation}
between $\bar{u}$ and $u$, with the latter defined in Eq.~\eqref{chem_devi_ug}. Rewriting Eq.~\eqref{chem_devi_ubar}, the solute concentration $c$ can be expressed as:
\begin{equation}
c=c^0_l(T) \frac{\left[ 1-g(\phi)+\delta \right]}{2} e^{\bar{u}}. \label{c_in_u}    
\end{equation}
By substituting $\phi = -1$ and Eq.~\eqref{u_u_bar} into Eq.~\eqref{c_in_u}, the solute concentration $c_l$ on the liquid side of the interface becomes:
\begin{equation}
c_l=c_{\infty} e^{u}. \label{measure_c_l}
\end{equation}
Thus, $c_l$ can be derived from $u$.

\begin{figure}[hbt!]
\includegraphics{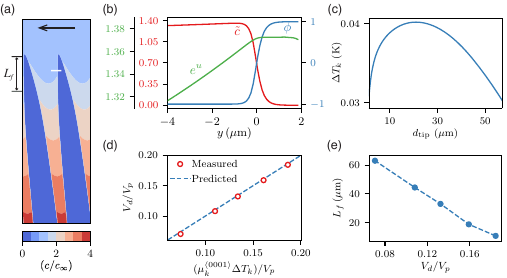}
\caption{\label{fig:figure5} (a) Colormap of the scaled solute concentration $\bar{c} \equiv c / c_{\infty}$ in a PF simulation with linear kinetics on the basal plane and $\mu_k^{\left<0001\right>} = 41.1 \, \mathrm{\mu m/s/K}$. The simulation domain size is $L_x \times L_y = 362 \times 120 \, \mathrm{\mu m^2}$, and $W_0 / d_0 = 2.5$. (b) Profiles of various fields across the faceted interface at the location indicated by the white line in (a). (c) Measured kinetic undercooling $\Delta T_k$ along the facet as a function of the distance below the tip, $d_{\mathrm{tip}}$. (d) Numerical measurements (dots) compared to the linear kinetic relation (dashed line). (e) Facet length $L_f$ as a function of the normalized drifting velocity, $V_d / V_p$.}
\end{figure}

As shown in Figs.~\ref{fig:figure5}(a)-(b), we obtain the $\phi$ and $c$ profiles across the diffuse interface in directions perpendicular to the facet that spans a distance $L_f$ below the tip of steady-state ice lamellae drifting at a velocity $V_d$. With the measured $\phi$ and $c$ profiles, we determine the $e^u$ profile using the form of $u$ defined in Eq.~\eqref{chem_devi_uh} in the nonvariational model. As shown in Fig.~\ref{fig:figure5}(b), the $e^u$ profile is nearly constant around the interface location ($y = 0$). The value of $c_l$ can then be accurately determined by substituting the measured $e^u$ value at the interface into Eq.~\eqref{sharp_interface_Tk}. 
With both $c_l$ and $T$, the kinetic undercooling $\Delta T_k$ is derived using Eq.~\eqref{sharp_interface_Tk}. We measure $\Delta T_k$ at different locations along the facet in PF simulations with linear kinetics on basal plane and plot $\Delta T_k$ as a function of the distance $d_{\mathrm{tip}}$ below the tip in Fig.~\ref{fig:figure5}(c). The maximum value of $\Delta T_k$ on the facet is then used to verify the kinetic relation for faceted growth. As shown in Fig.~\ref{fig:figure5}(d), the measurements indicate that the kinetic relation is well reproduced by the PF simulations, i.e., Eq.~\eqref{V_n_linear} with $V_n$ replaced by $V_d$. We verify this for five values of $\mu_k^{\left<0001\right>}$ ranging from $19.4$ to $96.9~\mathrm{\mu m/s/K}$ in different simulations. This agreement confirms that faceted growth is governed by interface kinetics and that the ice lamellae in PF simulations drift at velocities controlled by interface kinetics.
Additionally, we measure the facet length $L_f$ by defining a threshold for the end of the facet, $\tan(\theta) \leq 0.005$, where $\theta$ is the angle between the interface normal and the $c$-axis. As shown in Fig.~\ref{fig:figure5}(e), $L_f$ decreases with increasing $V_d$, indicating that the partially faceted structure strongly depends on basal plane kinetics.

\begin{figure}[hbt!]
\includegraphics[scale=0.6]{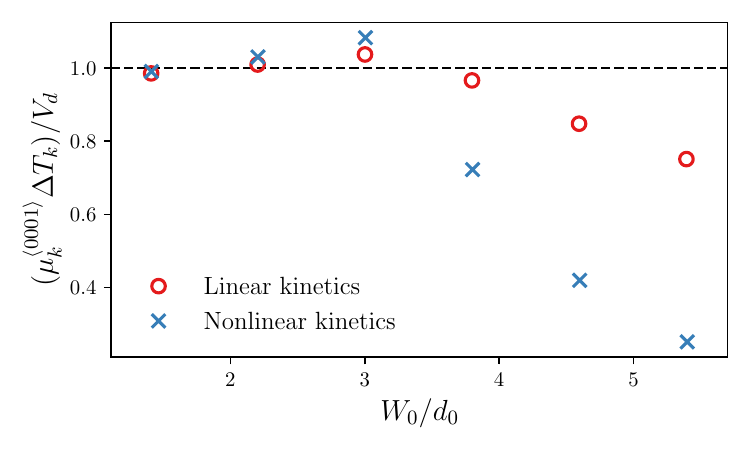}
\caption{Measured values of $(\mu_k^{\left<0001\right>} \Delta T_k)/V_d$ as a function of $W_0 / d_0$ for linear kinetics at $r = 32$ (circles) and nonlinear kinetics at $r_0 = 1.96$ (crosses) in 2D PF simulations.}  \label{Fig:convergence_kinetics}
\end{figure}

We also examine the effects of interface thickness on basal plane kinetics. In Sec.~\ref{sec:converge_Vd}, we showed that the drifting velocity $V_d$ converges sharply at a critical $W_c$. Here, we also measure $\Delta T_k$ and verify the kinetic relation for both linear and nonlinear kinetics at different $W_0$. As shown in Fig.~\ref{Fig:convergence_kinetics}, we plot $(\mu_k^{\left<0001\right>} \Delta T_k) / V_d$ as a function of interface thickness. This ratio equals $1$ when the drifting of ice lamellae follows the imposed kinetic relationship. For linear kinetics, $\mu_k^{\left<0001\right>}$ is constant, whereas for nonlinear kinetics, $\mu_k^{\left<0001\right>}$ is a function of $\Delta T_k$, as given in Eq.~\eqref{mu_k}. 
At larger interface thicknesses, $(\mu_k^{\left<0001\right>} \Delta T_k) / V_d \leq 1$. As $W_0 / d_0$ decreases to approximately $3$, the basal plane kinetics in simulations with both linear and nonlinear kinetics converges. Further decreasing $W_0 / d_0$ results in $(\mu_k^{\left<0001\right>} \Delta T_k) / V_d$ slowly approaching $1$, indicating that the drifting of ice lamellae becomes entirely kinetics-controlled.

\section{Conclusion and outlook}

In conclusion, we have presented a detailed PF model for microstructural pattern formation during ice templating. With the quantitative implementation of the anisotropic properties of both the excess interface free energy and atomic attachment kinetics, the proposed model accurately simulates a highly anisotropic ice-water interface. This interface is faceted in the basal plane normal to the $\left<0001\right>$ directions and atomically rough in other directions contained within the basal plane.
We used the PF model to simulate the directional solidification of a dilute binary aqueous solution within a temperature gradient $G$, where the $\left<11\bar{2}0\right>$ preferred growth direction of ice crystals is aligned with $G$. 3D simulations incorporating only highly anisotropic interface kinetics reproduce unilateral lamellar structures with a disordered rough side. While simulations including only weakly anisotropic interface free energy cannot independently induce the formation of partially faceted structures, they control the solidification front and select steady-state ice tips, which govern the formation of unilateral surface features on the rough side of the lamellae. The accurate implementation of both anisotropies is essential for quantitatively reproducing the hierarchical structure formation observed in experiments.

Furthermore, we showed that the drifting velocity $V_d$ of ice lamellae converges sharply at a critical interface thickness $W_c$. This convergence is influenced by the slope that determines the rate of variation of the kinetic coefficient with orientation. Once convergence is achieved, $V_d$ is solely controlled by the basal plane kinetics and is unaffected by the slope. Meanwhile, the tip radius and undercooling converge gradually as the interface thickness decreases, with the converged values being affected by the magnitude of the slope assumed in the form of the kinetic anisotropy. Although the shape of the kinetic anisotropy is not precisely known, the effective slope is estimated to lie within a broad range, where the upper bound presents significant computational challenges in PF simulations.
We have demonstrated that, with computationally tractable choices of the interface thickness and the shape of the kinetic anisotropy, the PF simulations remain accurate. This enables PF simulations at experimentally relevant length and time scales and the quantitative comparison with the ice-templated hierarchical structures formed in experiments.

This work can be extended in several directions. 
\added{Firstly, the anisotropic functions for interface free-energy and kinetics can be further improved based on the development of experimental and atomistic modeling studies of the ice-water interface during freezing.} \added{Secondly, the method for implementing highly anisotropic interfaces in PF modeling can be broadly applied to simulate ice-water and ice-vapor interfaces across a wide parameter space, as well as to other material systems that exhibit faceted growth. The current formulation focuses on ice-water interfaces in freeze casting, where interfaces within the basal plane are assumed to be atomically rough (i.e., above the roughening transition). By incorporating faceted directions within the basal plane, for example, the formulation can be extended to model ice-vapor interfaces below the roughening transition, where the driving forces are usually small. Meanwhile, by selecting appropriate anisotropy functions for interface free energy and kinetics, the PF model can be adapted to simulate faceted growth for other crystal structures.}
Thirdly, the PF model can be generalized for the directional solidification of colloidal suspensions, which are commonly encountered in freeze-casting experiments. In this context, more complex physical mechanisms than Brownian diffusion need to be considered for large-size particles that form a porous layer at the solidification front  \cite{Peppin2006SolidificationSuspensions} and can be engulfed by the advancing front \cite{Peppin2007MorphologicalSuspensions,Peppin2008ExperimentalSuspensions}.
Lastly, the PF model can be extended by coupling additional multiphysics processes and in particular the solutal flow arising from the volume expansion of ice that can significant affect the templating of structures in between ice lamellae. 

\section*{Acknowledgment}
This work was supported through NASA grants 80NSSC18K0305 and 80NSSC21K0039. The numerical simulations were performed on Northeastern University's Discovery cluster located in Massachusetts Green High Performance Computing Center (MGHPCC) in Holyoke, MA. Computing support for part of this work came from the LLNL Institutional Computing Grand Challenge program. K.J. acknowledges that this work was partially supported by Lawrence Livermore National Laboratory (LLNL) under Contract DE-AC52-07NA27344. The authors thank Ulrike Wegst for valuable discussions linked to experimental observations.

\appendix

\section{Phase-field model with finite partitioning} \label{sec:k-model}

The quantitative PF model for the solidification of binary alloys defines a finite partition coefficient $k = c_s / c_l$ \cite{Karma2001Phase-fieldSolidification,karma_phase-field_2003,Echebarria2004QuantitativeSolidification}, which can also be adapted to model ice templating in the limit $k \to 0$. In the conventional model, the reference temperature $T_0$ is defined as the temperature corresponding to $c_s = c_\infty$ on the solidus and $c_l = c_\infty / k$ on the liquidus. However, this reference temperature cannot be applied to the present model for ice templating, as $c_\infty / k$ diverges in the limit $k \to 0$. Therefore, we modify the reference temperature in this model to $T_0 = T_M - |m| c_\infty$. 
To differentiate the $\delta$-model with complete partitioning discussed in the main text, this PF model with finite partitioning is referred to as the $k$-model. Using the same free-energy and kinetic anisotropy functions as in the $\delta$-model, the evolution equations of the $k$-model are expressed as:
\begin{eqnarray}
&&\left[ \tilde{c}_l^0(T)+ A(\mathbf{n}) \right] a_s(\mathbf{n})^{2} \frac{\partial \phi}{\partial t} = \vec{\nabla} \cdot\left[a_s(\mathbf{n})^{2} \vec{\nabla} \phi\right] \nonumber \\
&&\qquad+\sum_{m}\left[\partial_{m}\left(|\vec{\nabla} \phi|^{2} a_s(\mathbf{n}) \frac{\partial a_s(\mathbf{n})}{\partial\left(\partial_{m} \phi\right)}\right)\right]+\phi-\phi^{3} \\
&&\qquad-\lambda (1-\phi^2)^2 \frac{1}{1-k} \left[ \frac{2\tilde{c}}{1+k-(1-k)\phi}-\tilde{c}_l^0(T) \right], \nonumber
\end{eqnarray}
\begin{eqnarray}
\frac{\partial \tilde{c}}{\partial t}=&& \widetilde{D} \vec{\nabla} \cdot \left( q(\phi) \tilde{c} \vec{\nabla} \ln \frac{\tilde{c}}{1+k-(1-k)\phi} \right) \nonumber\\
&&-\frac{1}{\sqrt{2}} \vec{\nabla} \cdot \left[ \tilde{c} \, \partial_t \ln (1+k-(1-k)\phi) \frac{\vec{\nabla} \phi}{|\vec{\nabla} \phi|} \right],
\end{eqnarray}
where the interpolation function for solute diffusivity across the diffuse interface has the form:
\begin{equation}
q(\phi)=\frac{1-\phi}{1+k-(1-k)\phi}, \label{interpolation_q}
\end{equation}
which reduces to 1 in the limit of $k \to 0$. The dimensionless quantities in these two equations have the same definitions as in the $\delta$-model.

Since the $k$-model is directly adapted from the quantitative PF model for alloy solidification, the only differences being the modified reference temperature and anisotropies, the asymptotic analysis presented in Ref.~\cite{Echebarria2004QuantitativeSolidification} also applies to the $k$-model. Hence, the numerical model still solves quantitatively the same sharp-interface equations.
Additionally, both the $k$-model and $\delta$-model are expected to yield similar quantitative results. In the liquid phase, the evolution equations of both models reduce to the identical form in the limits $k \to 0$ and $\delta \to 0$, respectively. In the solid phase, the two small parameters $k$ and $\delta$ play analogous roles in each model: they regularize the equations to prevent divergence as $\phi \to 1$. Thus, the differences between the models are primarily due to the specific regularization forms employed.
Consequently, the $k$-model, as used in Ref.~\cite{Yin2023HierarchicalTemplating}, and the $\delta$-model, as used in this paper and in Ref.~\cite{Ji2025ParityTemplating}, solve the same sharp-interface equations and produce similar faceted structures.
Either model can be chosen to simulate ice templating, with the $\delta$-model being slightly more efficient due to its use of one fewer interpolation function.

\added{The small but finite partitioning during ice crystal growth can also be achieved using a different PF approach~\cite{huang2025thermodynamically}, where a pseudo-component is introduced in the solid phase and its chemical potential is fitted to the phase diagram.}

\section{Liquidus slope} \label{Sec:phase_diagram}

In PF simulations, we use a linear analytical liquidus slope estimated by the Clausius-Clapeyron relation for dilute alloys \cite{Karma1993}, with $k \to 0$:
\begin{equation}
|m_v| = \frac{k_B T_M^2}{\Delta h_f (1-k)} \frac{1}{m_B}, \label{liquidus_m_v}
\end{equation}
where $m_v$ is the liquidus slope in units of Kelvin per weight by unit volume, $k_B$ is the Boltzmann constant, $T_M$ is the melting temperature of the pure solvent, $\Delta h_f$ is the latent heat of melting per unit volume, and $m_B$ is the molecular mass of the solute.

Since experimentally determined phase diagrams often express concentration in weight percent, we also derive an expression for the liquidus slope in units of Kelvin per weight percent. Ignoring the volume change due to mixing, Eq.~\eqref{liquidus_m_v} can be rewritten as:
\begin{equation}
|m| = \frac{k_B T_M^2}{\Delta h_f (1-k)} \frac{\rho_A \rho_B}{\left(1-c_w\right) m_B \rho_B + c_w m_B \rho_A},
\end{equation}
where $m$ is the liquidus slope in units of Kelvin per weight percent, $\rho_A$ is the density of the solvent, $\rho_B$ is the density of the solute, and $c_w$ is the weight fraction of the solute. 
In the dilute limit $c_w \ll 1$, the expression for $k \ll 1$ simplifies to:
\begin{equation}
|m| \approx \frac{k_B T_M^2 \rho_A}{\Delta h_f m_B}. \label{liquidus_dilute}
\end{equation}
With water as the solvent, where $\rho_A = 1~\mathrm{g/cm^3}$, we obtain $|m| \times 1~\mathrm{wt.\%} = |m_v| \times 1~\% \, \mathrm{w/v}$, where $\mathrm{w/v}$ denotes the weight by unit volume.
For sucrose and trehalose, both with a molar mass of $342.3~\mathrm{g/mol}$, the liquidus slope of their aqueous solutions in the dilute limit is $|m| \approx 0.0543~\mathrm{K/wt.\%}$. This analytically calculated liquidus slope has been shown in good agreement with experimental measurements \cite{Yin2023HierarchicalTemplating,ablett1992differential}.

\section{Numerical implementation of the phase-field model} \label{Sec:Dis_evo_eqs}

The model equations \eqref{dmodel_phi}–\eqref{dmodel_c} are solved on a square lattice in 2D and a cubic lattice in 3D using a finite difference method for spatial derivatives and an explicit Euler time-stepping scheme. For the leading differential terms, including the Laplacian and divergence, we use simple discretizations with a single set of lattice points \cite{ji2022isotropic}. For the anisotropy terms in Eq.~\eqref{dmodel_phi}, they are first analytically expanded into first- and second-order derivatives of $\phi$, and then solved on a regular stencil following the procedure outlined in Ref.~\cite{tourret2015growth}. 
Additionally, we provide details for several numerical techniques used here, including the discretization of the divergence and anti-trapping terms in Eq.~\eqref{dmodel_c}, and the accurate identification of the local angles $\theta$ and $\varphi$.

\subsubsection*{Discretization of the divergence term}

To numerically solve the diffusion term in Eq.~\eqref{dmodel_c}, we utilize the properties of the logarithmic terms to optimize spatial discretizations and enhance numerical stability. Consider a generalized form of the diffusion term:
\begin{equation}
\vec{\nabla} \cdot \vec{F}= \vec{\nabla} \cdot \left( \alpha\vec{\nabla} \ln{\beta} \right).
\end{equation}
In the $\delta$-model, the coefficients are defined as:
\begin{equation}
\alpha=\widetilde{D}\tilde{c} \quad \textrm{and} \quad \beta = \frac{\tilde{c}}{1-\phi+\delta}.
\end{equation}
The discretization of this term in 3D employs a simple set of six fluxes:
\begin{eqnarray}
\vec{\nabla} \cdot \vec{F} =&& \frac{1}{h} \left [ \vec{F}_{(1/2,0,0)}+\vec{F}_{(\bar{1}/2,0,0)}+\vec{F}_{(0,1/2,0)} \right. \nonumber \\
&&\left.+\vec{F}_{(0,\bar{1}/2),0}+\vec{F}_{(0,0,1/2)}+\vec{F}_{(0,0,\bar{1}/2)} \right], \label{dis_Flux}
\end{eqnarray}
where one of the fluxes is computed as:
\begin{equation}
\vec{F}_{(1/2,0,0)} = \frac{\alpha_{(1,0,0)}+\alpha_{(0,0,0)}}{2} \cdot \frac{\ln \left( \beta_{(1,0,0)}/\beta_{(0,0,0)} \right) }{\Delta x}. \label{divergence_F12}
\end{equation}
Similar discretizations for the other flux terms in Eq.~\eqref{dis_Flux} are obtained by applying translation and rotation operations. Since the neighboring values of $\beta$ are close to each other, ratios such as $\beta_{(1,0,0)} / \beta_{(0,0,0)}$ avoid extreme values, making the logarithmic term in Eq.~\eqref{divergence_F12} numerically more stable.

\subsubsection*{Discretization of the anti-trapping term}

The anti-trapping term in Eq.~\eqref{dmodel_c} can be written in a generalized form as:
\begin{equation}
\vec{\nabla} \cdot \vec{F}= \vec{\nabla} \cdot \left( \alpha \frac{\vec{\nabla} \phi}{|\vec{\nabla} \phi|} \right).
\end{equation}
In the $\delta$-model, the coefficient $\alpha$ is given by:
\begin{equation}
\alpha=-\frac{\tilde{c}}{\sqrt{2}} \partial_t \ln{(1-\phi+\delta)}.
\end{equation}
The discretization of $\vec{\nabla} \cdot \vec{F}$ follows the same approach as in Eq.~\eqref{dis_Flux}. One of the fluxes is computed as:
\begin{equation}
\vec{F}_{(1/2,0,0)} = \frac{\alpha_{(1,0,0)}+\alpha_{(0,0,0)}}{2} \cdot \frac{\phi_{(1,0,0)}-\phi_{(0,0,0)}}{h|\vec{\nabla} \phi|_{(1/2,0,0)}}. \label{F_1200}
\end{equation}
Since $\phi$ at the next time step has already been calculated at each lattice point after solving Eq.~\eqref{dmodel_phi}, it can be used to evaluate $\alpha$ at the current time step. Thus, we have:
\begin{equation}
\alpha=-\frac{\tilde{c}}{\sqrt{2} \Delta t} \ln{(\frac{1-\phi_{n+1}+\delta}{1-\phi_n+\delta})},
\end{equation}
where $\Delta t$ is the time step, and $\phi_n$ and $\phi_{n+1}$ are the values of $\phi$ at the current and next time steps, respectively. The gradient at the off-lattice point in Eq.~\eqref{F_1200} is evaluated as:
\begin{eqnarray}
&&|\vec{\nabla} \phi|_{(1/2,0,0)}=\\
&&\sqrt{\left( \partial_x \phi|_{(1/2,0,0)} \right)^2+\left( \partial_y \phi|_{(1/2,0,0)} \right)^2+\left( \partial_z \phi|_{(1/2,0,0)} \right)^2}, \nonumber
\end{eqnarray}
where 
\begin{equation}
\partial_x \phi|_{(1/2,0,0)}=\frac{\phi_{(1,0,0)}-\phi_{(0,0,0)}}{h},
\end{equation}
\begin{equation}
\partial_y \phi|_{(1/2,0,0)}=\frac{\phi_{(1,1,0)}+\phi_{(0,1,0)}-\phi_{(1,\bar{1},0)}-\phi_{(0,\bar{1},0)}}{4h},
\end{equation}
\begin{equation}
\partial_z \phi|_{(1/2,0,0)}=\frac{\phi_{(1,0,1)}+\phi_{(0,0,1)}-\phi_{(1,0,\bar{1})}-\phi_{(0,0,\bar{1})}}{4h}.
\end{equation}
Similar discretizations for the other flux terms in Eq.~\eqref{dis_Flux} can be applied through translation and rotation operations.

\subsubsection*{Accurate identification of the interface orientation} \label{Sec:Dis_orientation}

With strong kinetic anisotropy, the relaxation time $\tau(\mathbf{n})$ can vary rapidly near the $\left<0001\right>$ directions and is highly sensitive to the angle $\theta$. Therefore, accurately identifying the interface orientation with a finite-difference scheme is crucial in PF simulations. Similar to the isotropic discretization of differential terms \cite{ji2022isotropic}, we use two sets of lattice points to determine the interface normal directions. In 2D, the orientation is given by:
\begin{equation}
\sin\theta = \frac{\phi_x}{\sqrt{\phi_x^2+\phi_y^2}} ,\, \cos\theta = \frac{\phi_y}{\sqrt{\phi_x^2+\phi_y^2}}. \label{sin_cos}
\end{equation}
The derivatives $\phi_x$ and $\phi_y$ can be discretized using lattice points in the $\left<10\right>$ directions, i.e., $\phi_x = \phi_{10}$ and $\phi_y = \phi_{01}$, where $\phi_{10}$ and $\phi_{01}$ represent derivatives of the $\phi$ field in the $[10]$ and $[01]$ directions, respectively. Alternatively, the lattice points in the $\left<11\right>$ directions can be used to independently evaluate $\theta$, with $\phi_x=(\phi_{11}+\phi_{1\bar{1}})/\sqrt{2}$ and $\phi_y=(\phi_{11}+\phi_{\bar{1}1})/\sqrt{2}$. Each set of lattice points is used to independently compute $\sin\theta$ and $\cos\theta$, and their final values are obtained by weighting the $\left<10\right>$ lattice with $2/3$ and the $\left<11\right>$ lattice with $1/3$.

In 3D, the angles $\theta$ and $\varphi$ in Eq.~\eqref{surf_energy_anis} are given by:
\begin{equation}
\sin\theta = \frac{\sqrt{\phi_x^2+\phi_y^2}}{\sqrt{\phi_x^2+\phi_y^2+\phi_z^2}} ,\, \cos\theta = - \frac{\phi_z}{\sqrt{\phi_x^2+\phi_y^2+\phi_z^2}},
\end{equation}
and
\begin{equation}
\sin\varphi = -\frac{\phi_y}{\sqrt{\phi_x^2+\phi_y^2}} ,\, \cos\varphi = -\frac{\phi_x}{\sqrt{\phi_x^2+\phi_y^2}}.
\end{equation}
Similarly, we use lattice points in the $\left<100\right>$ and $\left<110\right>$ directions to evaluate the interface orientation. For the $\left<100\right>$ lattice points, $\phi_x=\phi_{100}$, $\phi_y=\phi_{010}$ and $\phi_z=\phi_{001}$. For the $\left<110\right>$ lattice points, $\phi_x=\sqrt{2}(\phi_{110}+\phi_{101}+\phi_{1\bar{1}0}+\phi_{10\bar{1}})/4$, $\phi_y=\sqrt{2}(\phi_{110}+\phi_{011}+\phi_{\bar{1}10}+\phi_{01\bar{1}})/4$ and $\phi_z=\sqrt{2}(\phi_{101}+\phi_{011}+\phi_{\bar{1}01}+\phi_{0\bar{1}1})/4$. The trigonometric functions are computed independently for each lattice, and the final values are obtained by weighting the $\left<100\right>$ lattice with $1/3$ and the $\left<110\right>$ lattice with $2/3$.

\bibliography{apssamp}

\end{document}